\pdfoutput=1
\documentclass{article}
\usepackage[table]{xcolor}
\usepackage[utf8]{inputenc}
\usepackage[margin=1in]{geometry} 

\usepackage[sorting=none]{biblatex}
\usepackage{nameref,hyperref}
\RequirePackage{tikz,wrapfig}
\usetikzlibrary{automata,arrows,positioning,shapes,calc}

\usepackage{array}

\RequirePackage{amsthm,amsmath,amsfonts,amssymb}
\RequirePackage{graphicx} 
\RequirePackage{mathtools,wrapfig}
\RequirePackage{caption,subcaption}
\mathtoolsset{showonlyrefs}

\theoremstyle{plain}

\theoremstyle{definition}

\theoremstyle{remark}

\DeclareSymbolFont{bbold}{U}{bbold}{m}{n}
\DeclareSymbolFontAlphabet{\mathbbold}{bbold}
\def\one{\mathbbold{1}}

\usepackage{setspace} 
\begin{document}

\begin{flushleft}
{\Large
\textbf\newline{Bayesian Forensic DNA Mixture Deconvolution Using a Novel String Similarity Measure}
}
\newline
\\
Taylor Petty\textsuperscript{1*\textcurrency},
Jan Hannig\textsuperscript{1,2},
Hari Iyer\textsuperscript{2}
\\
\bigskip
\textbf{1} Statistics \& Operations Research, University of North Carolina at Chapel Hill, Chapel Hill, NC, USA
\\
\textbf{2} Statistical Engineering Division, Information Technology Laboratory, National Institute of Standards and Technology, Gaithersburg, MD, USA
\\
\bigskip

\textcurrency Current Institution: Sciome LLC, Research Triangle Park, NC, USA 

* taylor.michael.petty@gmail.com

\end{flushleft}

\section*{Abstract}
Mixture interpretation is a central challenge in forensic science, where evidence often contains contributions from multiple sources. In the context of DNA analysis, biological samples recovered from crime scenes may include genetic material from several individuals, necessitating robust statistical tools to assess whether a specific person of interest (POI) is among the contributors. Methods based on capillary electrophoresis (CE)
are currently in use worldwide, but offer limited resolution in
complex mixtures. Advancements in massively parallel sequencing (MPS)
technologies provide a richer, more detailed representation of DNA
mixtures, but require new analytical strategies to fully leverage this
information. In this work, we present a Bayesian framework for evaluating whether a POI’s DNA is present in an MPS-based forensic sample. The model accommodates known contributors, such as the victim, and uses a novel string edit distance to quantify similarity between observed alleles and sequencing artifacts. The resulting Bayes factors enable effective discrimination between samples that do and do not contain the POI’s DNA, demonstrating strong performance in both hypothesis testing and classification settings.

\section{ Introduction}\label{sec:intro}

In forensic science, mixture interpretation is a fundamental challenge encountered across a wide range of evidence types. Complex forensic samples often consist of overlapping contributions from multiple sources, requiring deconvolution techniques to disentangle and identify individual components. In drug identification, for example, seized materials may contain a combination of illicit substances and cutting agents; analytical methods such as mass spectrometry necessitate deconvolution to accurately determine the presence or absence of specific compounds. Similarly, in forensic audio analysis, recordings of overlapping voices require source separation techniques to ascertain whether a particular individual participated in a conversation. Trace chemical analysis, environmental forensics, and digital signal investigations frequently rely on analogous strategies where mixture resolution is critical to evidentiary interpretation. Among these applications, deconvolution plays an especially pivotal role in the interpretation of DNA mixtures. When biological material recovered from a crime scene contains genetic material from multiple contributors, forensic analysts must determine whether a specific individual, such as a suspect, is included among the contributors. Currently, such evaluations are based on capillary electrophoresis (CE) profiles generated after PCR amplification of short tandem repeat (STR) markers. This CE technology and methodology have been in use for decades. However, recent advances enable the sequencing of forensic DNA mixtures, offering a more detailed view of the constituent genetic material by counting sequence fragment types.

This paper is concerned with evidence analysis of mixtures of biological material from which forensic DNA is extracted to support or refute the claim that a given person of interest (POI) with a known DNA profile contributed DNA to a sample at a crime scene. A primary goal is to give a numerical measurement comparing two models (hypotheses) sharing a common number of contributors $k$:
\begin{enumerate}
    \item A model where the POI contributed to the mixture, which hypothesis will be referred to as $M_1$, also known as the prosecution hypothesis.
    \item A model where the mixture contains DNA from random members of the population unrelated to the POI. This hypothesis will be referred to as $M_2$, also known as the defense hypothesis.
\end{enumerate}

 A random member of the population may match the POI's genetic profile, although the chance that this happens is extremely small \cite{GettingsKieslerVallone2015, Panneerchelvam2003-vg}. Additionally, in certain situations (e.g., sexual assault) DNA from one or more individuals, other than the POI, is expected to be present in the mixture. In such cases, these contributors may be constrained to be present in both models.
It is often assumed that the unknown contributors are unrelated to each other and to the known contributors. If the prosecution or defense claims relatedness among the contributors, separate models must be employed to account for the dependence among the genotypes  that will be present under those models. These will not be discussed here, but the reader is referred to \cite{DNArelatedcontributors_kruijver2021}.

The chosen method for comparing the models $M_1$ and $M_2$ is the Bayes factor. A Bayes factor is the ratio $B_1/B_2$, where $B_i$ is the marginal likelihood of $M_i$ averaged by integrating over its parameter space $\Theta_i$, also known as the marginal likelihood: \[\frac{B_1}{B_2} = \frac{Pr(\Lambda\mid M_1)}{Pr(\Lambda\mid M_2)} = \frac{\int_{\Theta_1} Pr(\theta_1\mid M_1) Pr(\Lambda\mid \theta_1, M_1)\,d\theta_1}{\int_{\Theta_2} Pr(\theta_2\mid M_2) Pr(\Lambda\mid \theta_2, M_2)\,d\theta_2}.\]

There are two common ways of using Bayes factors. The first is in a Bayesian decision framework, where Bayes factors are used to classify whether a POI is present in the mixture or not based on whether the Bayes factor is larger or smaller then a threshold that is determined by error costs and prior probabilities.
Another view of the Bayes factor is as an indication of how many times more likely the observed data is under $M_1$ than under $M_2$. In that light it becomes a relative weight of evidence measure, a numerical score weighting one model versus another.
 For more on Bayesian decision theory in forensics, see \cite{BIEDERMANN2018}. A brief introduction is also included in the Supplementary Material.

\subsection{ Forensic DNA identification}\label{intro:sec:background}

The most common form of forensic DNA identification uses short tandem repeat (STR) regions of the genome \cite{variationinstr_1994, rapid_efficient_1994}. STR regions vary substantially across the population, and barring identical twins it is vanishingly unlikely for two people to match at many STR locations. Thus, STR sequences have very high discriminating power, being both accurate and sensitive \cite{dnatyping_str_1993}.

The forensic community has selected specific regions across the genome known as loci (singular: locus) where STRs are present. At a particular locus, all individuals have a sequence that follows the general form [\emph{motif}]\emph{k}, where \emph{motif} is  usually 3-5 nucleotides long, and $k$ is the number of consecutive repeats of that motif. One often describes this by saying that the individual possesses a $k$-allele at that locus, or has an allele $k$ at the locus. This repeating region is surrounded by a non-repeating section of DNA called a flanking region \cite{ngsoverview}. One of the main differences between sequencing kits is the resulting flanking region in the product sequences, but the targeted STR region remains the same and the model described in this paper can be adapted to different kits. 

At each locus an individual has two allelic sequences, one allele inherited from each parent. These two alleles could have the same sequence (homozygous), e.g., an allele with 8 STR repeats (8-allele for short)  from both parents, or different types (heterozygous), e.g., a 7-allele from the mother and a 10-allele from the father. 

DNA samples recovered from crime scenes are often DNA mixtures from two or more individuals contributing at different proportions. This results in the need to deconvolve the mixture to infer the genotypes of individual contributors.
Typically DNA recovered from a crime sample is amplified using the polymerase chain reaction (PCR). This process introduces multi-nucleotide errors called stutters -- the addition or deletion of an entire STR motif at once -- as well as errors due to insertion, deletion, and substitution of single nucleotides. These error sequences will be collectively referred to as artifacts. For details, refer to \cite{characterizingstutter_BROOKES2012, FDSTools_HOOGENBOOM2017, characterizeMPSSTRstutter_AGUDO2022}.

\subsection{ Data processing and output}

After PCR the sample is sequenced, then analyzed using bioinformatics algorithms. Finally, a table of sequences and their associated counts is produced for each locus. Since PCR errors are generated from the alleles originally present in the sample, we call these original alleles {\em parent sequences}, borrowing from graph theory terminology to reflect their nature as the root in the process leading to the observed artifact sequences. In typing a standard single-source sample for a set of autosomal loci (i.e., non-lineage markers), the sequences observed should include one to two parent alleles, depending on whether the sample was homozygous or heterozygous at the autosomal locus, respectively. Additionally, numerous artifact sequences are present because of copying errors in the amplification and sequencing process. Since these artifact sequences originated as erroneous copies of true parent alleles, they will hereafter be referred to as {\em child sequences}. The number of times each sequence is observed is commonly referred to as the depth of coverage (DoC). An example of sequencing data is shown in Table~\ref{toydata_het}. 
The sequences with highest DoC represent the parental alleles and all the others are PCR or sequencing artifacts (i.e., child sequences). 

\begin{table}
\centering
\begin{tabular}{ |l|l|l| }
 \hline
 Sequence & DoC & Artifact Type \\ \hline
 TTAG [AACG]22 GGTCA & 2092 & Allele 1\\
 AC [AACG]6 TCCG & 1955 & Allele 2\\
 AC [AACG]\textcolor{red}{5} TCCG & 172 & Back stutter of allele 2\\
 TTAG [AACG]\textcolor{red}{21} GGTCA & 156 & Back stutter of allele 1\\
 TTAG [AACG]\textcolor{red}{20} GGTCA & 9 & Double back stutter of allele 1\\
 AC [AACG]6 TCCG\textcolor{red}{G} & 1 & 1 base pair insertion in allele 2\\
 \vdots & \vdots & \vdots\\
 \hline
\end{tabular}
\caption[Example heterozygous data.]{A synthetic example of an amplified and sequenced marker from one person. This person is heterozygous at this locus, as shown by the two different allelic sequences (Alleles 1 and 2) with highest DOC (top two rows). The first two rows are true alleles from the person's DNA, and all other sequences are errors generated during PCR amplification and sequencing.
}\label{toydata_het}
\end{table}

As sequencing technology has developed, individual nucleotides within STR regions in the human genome can be measured with increasing accuracy and efficiency,\newline whereas in the past scientists were limited to measuring the total length of the targeted sequence.  Consider an example of a locus with two alleles, with repeat regions A = [TCTA]8 [TCTG]1 [TCTA]1 and B = [TCTA]10. A and B are identical by length but different in sequence. These two alleles will be classified as homozygous (10, 10) if the sample is genotyped using the traditional (and still in-use) capillary electrophoresis (CE) method 
\cite{strcurrentfuture, NGSdiscrimbetterthanCE_Alvarez2017, NGSdiscrimbetterthanCE_Rockenbauer2014}. By recognizing intra-motif variations at the sequence level with modern sequencing technology, the power to discriminate between individuals increases \cite{nist2016paper}. New algorithms are required to handle this new data type, and that is what this paper addresses.

Consider the complexity that arises in data like Table~\ref{toydata_het} when multiple contributors are present, e.g., two contributors at 90 \% -- 10 \% proportions. Artifact noise from the 90 \% contributor can easily overlap with the allele signal of the 10 \% contributor, greatly complicating the deconvolution. Additionally, in the case of two contributors the number of true alleles can vary from one to four. If both contributors are homozygous at a locus by sequence, there will be one true allelic sequence; if both are heterozygous, there will be either four true distinct allelic sequences or two or three sequences depending on the degree of allele sharing between the contributors. 
With three contributors the number of possibilities range from one to six, and so forth with more contributors. 

The FBI has a standard suite of 20 loci named the Combined DNA Index System Core Loci, also known as the CODIS 20 \cite{codisinfo,NISTCODIS_1997,FBICODIS}. For this work, the Penta D and Penta E loci were included along with the CODIS 20 due to the interest of the collaborating geneticists at NIST, building on the work in \cite{confproceedings_whypentaDE1999Bacher}. Thus, in this paper, the data used for modeling were full allelic sequences at 22 autosomal STRs (20 CODIS core loci in addition to Penta D and Penta E), provided by the Applied Genetics Group at NIST. There were 661 single-source DNA samples collected from four different U.S. populations. They were sequenced using the PowerSeq Auto/Y System kit (Promega Corp., Madison, WI, USA) and analyzed using STRait Razor v3.0 \cite{previousSTRanalysis}.

\subsection{ String edit distances}\label{intro:sec:stringeditdists}

Modeling the distribution of PCR artifacts is facilitated by a characterization of how similar or different any two sequences are. String edit distances, based on counting the operations it takes to change one string into another, are one way to accomplish this.

Most string edit distances consider only single character edits and are variations of Levenshtein distance. These standard string edit distances are inadequate for analyzing similarity between DNA sequences due to their inability to separately address whole-motif edits. To fill this gap, a new string edit distance was defined, referred to as Restricted Forensic Levenshtein (RFL) distance, that accommodates the addition or deletion of one or more motifs as edit types separate from single-nucleotide insertion, deletion, and substitution. 
The background and details of the RFL distance can be found in \cite{petty2022}. The code is available on GitHub \cite{mygithub}.

\subsection{ Outline}
The remainder of this paper is organized as follows. Section~\ref{sec:model} discusses the likelihood, priors, estimator, MCMC, and other details of the model. Section~\ref{sec:sim} reviews the computational results. Sections~\ref{sec:discussion} and \ref{sec:conclusion} discuss the findings and provide concluding remarks.

\section{ Bayes factor for forensic identification}\label{sec:model}

This section outlines the construction of a Bayes factor for comparing the prosecution ($M_1$) and defense ($M_2$) hypotheses described in Section~\ref{sec:intro}. 
In Section~\ref{model:sec:biglik}, a mixture likelihood is formulated, with priors set in Section~\ref{model:sec:priors}. Priors, likelihood, and marginalization are all combined to give a single joint density in Section~\ref{model:subsec:marginalizingnuisance}. The Bayes factor estimation is discussed in Section~\ref{model:sec:bayesfac}. In the Supplementary Material, training data is used to find probabilities of each edit type, and these probabilities are used to obtain edit costs and other parameters needed in the model.

\subsection{ Model}\label{model:sec:biglik}

A forensic DNA mixture is comprised of $k$ contributing persons for a fixed $k\geq 2$. For the work presented here, each person is sequenced at $L=22$ loci, with specifics on loci chosen and the motifs analyzed for those loci shown in Table 2 of \cite{petty2022}. Determining what $k$ should be from data alone is a separate question that will not be addressed here, but the reader is referred to \cite{Paoletti2012_inferringnumbercontributors}. The process in this paper can be run for various $k$ and results can be compared. 

At each locus $l$, a list of possible alleles is prespecified from established population data. While CE has known alleles and their frequencies on the basis of length, full lists for MPS are still in development \cite{STRidER_Bodner2016, MPSSTR_nomenclature_Parson2016}. We denote this list of known alleles $K^l := [s^l_1,\dots,s^l_{J^l}]$, where $J^l := |K^l|$. The original DNA sample contains proportions $r^l$ of parent alleles at each locus $l$, i.e., at locus $l$ the $j$-th allele of $K^l$ is present at proportion $r^l_j$. Note that many of the $r^l_j$ may be $0$.
The number of artifact sequences in the data at locus $l$ is denoted by $M_l$.

We will model the observed data with the help of the following generative model. An observation is modeled to be generated independently for every locus $l=1,\dots,L$ as follows: 
\begin{enumerate}
    \item Randomly generate one of the predetermined possible parent sequences $s\in K^l$ as Categorical$(r^l)$, reflecting that alleles present in higher proportion are more likely to generate artifacts. This sequence $s$ is the parent that this single generated artifact is being generated from, reflecting that the artifacts in PCR amplification and MPS reads originate from true alleles.
    \item Generate a random distance $d$ according to a probability density $f$. 
    \item There may be several distinct artifact sequences at distance $d$ from the parent allele $k$. Select one of these uniformly at random to obtain the observed sequence $\lambda$. 
\end{enumerate}
This data-generating algorithm is repeated $M_l$ times and implies the following model likelihood:
\begin{equation}\label{eq:likelihood}
    \pi(D,j,\Lambda\mid r) = \prod_{l=1}^L \prod_{m=1}^{M_l} \frac{r^l_{j^l(m)} f(d^l_{j^l(m), m})}{C^l_{j^l(m)}(d^l_{j^l(m), m})},
\end{equation}
where $L$ is the number of loci, $M_l$ is the number of sequences at locus $l$, and $r^l_j$ is the proportion of parent sequences $s^l_j\in K^l$ at locus $l$ in the original sample. The function $j^l(m)$ intakes an artifact $m$ at locus $l$ and outputs the parent sequence in $K^l$ that artifact $m$ originated from. Note that $j^l(m)$ is a latent variable that will be marginalized out in the Bayes factor estimation procedure. The distance from allele $j^l(m)$ to artifact $m$ is $d^l_{j^l(m), m}$ and the corresponding likelihood of observing a sequence at that distance is $f(d^l_{j^l(m), m})$. The denominator $C^l_{k}(d)$ is the number of distinct potential artifact sequences that are exactly distance $d$ from parent allele $k$ at locus $l$. See the Supplementary Material for more details.

\subsection{ Priors}\label{model:sec:priors}

In Section~\ref{model:sec:biglik}, the vector of allele proportions $r^l$ was given. The prior for this parameter will be described in this section. A broad outline is given here, with additional details in the supplementary document. We need to set priors for the following parameters:
\begin{enumerate}
    \item Allele mixing proportions $p$, with $p_i$ representing the fraction of DNA material contributed by person $i$. 
    \item The contributor's profiles comprising of two (not necessarily unique) alleles $s_{l,j_1}$ and $s_{l,j_2}$ for each person $1,\dots,k$ and locus $1,\dots,l$ stored in an assignment matrix $A^l$. 
    \item The proportions of each parent allele $r^l$ for each locus.
\end{enumerate}

We select the non-informative prior on the proportions people contributed to the sample as $\pi(p)\sim$ Dir$(1/k,\dots,1/k)$, reflecting the prior belief that nobody is prone to be a higher-proportion contributor than anyone else. 

The proportion $p$ and the parent allele assignments are independent from each other since mixture proportions do not affect people's alleles.
Within one person, zygosity and alleles are assumed to be independent across loci. Note that the FBI chose the CODIS 20 loci so that independence would hold \cite{independentloci_Ge2012}).  For each locus $l = 1, \ldots, L$ define a $k\times J^l$ assignment matrix comprised of entries from $\{0,1/2,1\}:$
\begin{equation}
    A^l\in M(\{0,0.5,1\})^{k\times J^l},
\end{equation}
where the row sums of $A^l$ are constrained to equal 1, reflecting the possibility of homo- or heterozygous contributors. The rows of $A^l$ represent contributors $i=1,\dots,k,$ and the columns represent candidate alleles from the list $K^l$.
Thus the prior on $A_l$ is i.i.d across rows, reflecting the fact that we assumed that people (i.e. rows of $A^l$) in our mixture are random members of the population. 
The prior on each row 
$
\pi^l(A^l_{i\cdot})
$ 
should be chosen to reflect our knowledge of genotype frequencies in the reference population. See the Supplementary Material for more details.

Denote by $q^l=A^{l\top} p$ the theoretical proportion of each parent allele in the sample.  Surprisingly, the artifact assignment $r^l\sim$ Categorical$(q^l)$ does not exhibit sufficient variability compared to what we see in the data, and modeling $r^l\sim$ Dir$(c q^l)$ allows the proportions to vary more naturally, reflecting the inherent nature of the sequencing technology. Finally, for the coefficient $c$ we assume a log-normal prior with the hyperparameters chosen so that the prior has mean 22 and variance 3.
These values may need to be tuned if a different sequencing kit were to be used.

Thus the prior is
\begin{equation}\label{eq:prior}
\pi(p,c,A,r) = 
\pi(p)\pi(c)\prod_{l=1}^L 
\left(\pi^l(A^l) \pi^l(r^l\mid p,c,A^l)   \right),
\end{equation}
where $\pi^l(A^l)=\prod_{i=1}^k \pi^l(A^l_{i\cdot})$ with $A^l_{i\cdot}$ denoting $i$th row of matrix $A^l$, and 
\begin{gather*}
\pi(p) = \frac{\prod_{i=1}^k p_i^{1/k-1}}{Z_{Dir}(1/k)},\quad \mbox{with}\quad Z_{Dir}(\zeta) = \frac{\prod_{i=1}^K\Gamma(\zeta_i)}{\Gamma(\sum_{i=1}^K \zeta_i)};\\
\pi^l(r^l\mid p,c,A^l)=
\frac{\prod_{i=1}^{J^l} (r^l_i)^{c q^l_i-1}}{Z_{Dir}(c q^l)},\quad\mbox{where}\quad 
q^l = A^{l\top} p;\\
\mbox{and } \pi(c) = \frac{\exp\left(-(\ln(c)-\mu)^2/(2\sigma^2)\right)}{c\sigma\sqrt{2\pi}},\\ \quad
\mbox{with}\quad
\mu=\ln\left(\frac{22^2}{\sqrt{22^2+3}}\right)\mbox{ and  } 
\sigma^2=\ln\left(1+\frac{3}{22^2}\right).
\end{gather*}

\subsubsection{A Note on Dirichlet Parametrization}

The model will require Laplace approximation over a Dirichlet density, which requires an optimization step. The standard Dirichlet likelihood over values $x$ in the unit simplex has parameters $u_i>0$, but contains powers $u_i-1$. For $x_i\approx 0$ and $u_i<1$, the likelihood will contain a factor approaching infinity. This instability causes extreme difficulty for an optimization scheme when allele proportions are small, but the model should function well near these boundaries since it will almost certainly be testing alleles that are not present in the mixture (i.e., where the true Dirichlet parameter is near $0$).

Transforming parameters from the unit simplex to a multinomial logistic function overcomes this problem. Exposition is available in \cite{softmaxdirichlet_mackay1998}, but an alternative, more streamlined transformation is shown in the Supplementary Material that computes a relevant normalizing constant exactly instead of by approximation as MacKay does. This involves the introduction of a density $h_q$ over a free variable, which is defined as normal with mean 0 and variance equal to the number of non-zero entries of $q$. 

Recall that $q^l=A^{l\top} p$. We define $q_{\neq 0}^l$ to be the vector retaining only non-zero entries of $q^l$.
Thus $r^l$ are replaced in the likelihood \eqref{eq:likelihood} with $\mathbf{p}(a^l)=\frac{(e^{a_1},\dots,e^{a_{I}})}{\sum_{i=1}^{I} e^{a_i}}$, and the prior terms $\pi^l(r^l\mid p,c,A^l)$ in \eqref{eq:prior} are replaced by
\[
\begin{aligned}
\pi^l(a^l\mid p,c,A^l) &=
\frac{h_{q^l}(\sum a^l)\prod_{\{t : q_t^l\neq 0\}} \mathbf{p}(a^l)_t^{c q^l_t}}{Z_{Dir}(cq^l_{\neq 0})/|q^l_{\neq 0}|}, \\ 
\mbox{where}
\quad
h_q(a)&= \frac{\exp(-a^2/(2|q_{\neq 0}|))}{\sqrt{2\pi|q_{\neq 0}|}}.
\end{aligned}
\]

\subsection{ Marginalizing out nuisance parameters}\label{model:subsec:marginalizingnuisance}

The variables $j^l$ and $a^l$, l=1\ldots, L, are nuisance and therefore need to be marginalized out. The full details of the marginalization can be found in the Supplementary Material. For convenience, define
\[G^l(a^l) := \left(\frac{h_{q^l}(\sum a^l)\prod_{\{t : q_t^l\neq 0\}} \mathbf{p}(a^l)_t^{c q^l_t}}{Z_{Dir}(cq^l_{\neq 0})/|q^l_{\neq 0}|}\right)\prod_{m=1}^{M_l} \sum_{j=1}^{J^l}\frac{\mathbf{p}(a^l)_j f(d^l_{j,m})}{C^{l}_{j}(d^l_{j,m})},\]
with the $A^l,p,c,D,$ and $\Lambda$ implied by context for brevity. Define $R^l:=\mathbb{R}^{|q^l_{\neq 0}|}$ so that after marginalizing the joint distribution of data and parameters can be written as 
\begin{equation}
\pi(p,A,c,D,\Lambda) = \pi(p)\pi(c) \prod_{l=1}^L \left( \pi^l(A^l)\int_{R_l} G^l(a)\, da\right).\label{model:eqn:expandpriors}
\end{equation}

The Laplace approximation is used for computation of the integral (see, e.g., \cite{softmaxdirichlet_mackay1998}), where $\hat{a}$ denotes the argmax of a function $G^l$ over its domain and $\mathbf{H}^l(\cdot)$ denotes the Hessian of $\ln G^l$:
\begin{equation}\label{eq:LaplaceApp}
    \int G^l(x)\,dx \approx G^l(\hat{a}^l)(2\pi)^{d/2}|-\mathbf{H}^l(\hat{a}^l)|^{-1/2}.
\end{equation}
The \texttt{scipy.optimize.minimize} function from SciPy 1.8.1 was used with the Nelder-Mead method, max iterations 80,000, initialized to $\ln (cq_{\neq 0})$, unconstrained due to the multinomial logistic parametrization of the Dirichlet. We set the Laplace approximation to zero if the optimization procedure fails to converge, which happened extremely rarely. 

The marginal density \eqref{model:eqn:expandpriors} can now be used to derive the unnormalized conditional distribution $\pi(p,A,c \mid D,\Lambda)$ for use in Bayes factor estimation via a Markov chain Monte Carlo.

\subsection{ Model comparison: Bayes factor}\label{model:sec:bayesfac}

As discussed in Section~\ref{sec:intro}, the goal of this work is to obtain a Bayes factor for detecting a POI at a crime scene. Importance sampling is a common estimation method for Bayes factors from MCMC samples. We will be using a particular form available in the case of nested parameter spaces  \cite{MCmethodsBayes_Ibrahim2000}. Specifically, for densities $\pi_i=Q_i/B_i$ defined on $\Omega_i$ $i=1,2$, with $\Omega_1\subset\Omega_2$. Note that taking the expected value with respect to $\pi_2$ gives the following:
\begin{equation}\label{eq:BayesFactorIS}
    E_2\left[\frac{Q_1(\theta)}{Q_2(\theta)}\right] = \int_{\Omega_2} \frac{Q_1(\theta)}{Q_2(\theta)} \frac{Q_2(\theta)}{B_2}\,d\theta = \frac{1}{B_2} \int_{\Omega_1} Q_1(\theta)\,d\theta = \frac{B_1}{B_2}.
\end{equation}
Therefore the ratio of normalizing constants can be estimated as $$\frac{1}{N}\sum_{n=1}^N \frac{Q_1(\theta_{2,n})}{Q_2(\theta_{2,n})},$$
where $\theta_{2,n}$ are realizations of an ergodic Markov Chain with stationary distribution $\pi_2(\theta).$

Constraining the POI to be present is equivalent to constraining the space of matrices $A$ to contain the POI's profile at each $l$ as one of its rows. Let $\Omega_2=\bigotimes_{l=1}^L \Omega_2^l$, where $\Omega_2^l$ denotes the set of all matrices $A^l$  considered under hypotheses $M_2$ at loci $l$.  Define the sets $\Omega^{i,l}_1$ for $i\in\{1,\dots,k\}, l\in\{1,\dots,L\}$ as matrices $A^l\in\Omega_2^l$ matching the POI's genotype $s_l$ at row $i$ for locus $l$.
Set  $\Omega^i_1:=\bigotimes_{l=1}^L \Omega^{i,l}_1$, and 
$\Omega_1=\bigcup_{i=1}^k \Omega^i_1$. Set $\Omega_1$ represents the event that the POI's genotype is one of the genotypes contributing to the mixture. 

The desired Bayes factor is computed using \eqref{eq:BayesFactorIS} with $$Q_2(A,p,c)=\pi(p,A,c,D,\Lambda)I_{\Omega_2}(A),$$ and after adjusting the prior on $A$ for the fact that one of its row is known to be POI's genotype \[Q_1(A,p,c)=\frac{\pi(p,A,c,D,\Lambda)}{\prod_{k=1}^l\pi^l(S^l)} I_{\Omega_1}(A),\] where $S^l$ indicates the genotype of the POI at locus $l$ (i.e. the particular row of $A^l$ matching the POI at $l$). 
From here,
\begin{align}
    \frac{B_1}{B_2} &= {E_2}\left[E_2\left[\frac{Q_1(A,p,c)}{Q_2(A,p,c)} \mid p,c\right]\ \right]\\
    &\approx \frac{1}{N}\sum_{n=1}^N\sum_{i=1}^k\prod_{l=1}^L \sum_{A^l\in\Omega_2^l} \frac{I(A^l\in\Omega_1^{i,l})}{\pi^l(S^l)} \pi(A^l\mid p_n, c_n, D,\Lambda),\label{model:eqn:bayfaccalcsplit}
\end{align}
where $p_n$, $c_n$ are samples from an ergodic MCMC with the posterior $\pi(p,A,c \mid D,\Lambda)$ as its stationary distribution, and using
 $\pi(\Omega_1\mid c,p,D,\Lambda)\approx \sum_{i=1}^k \pi(\Omega^i_1\mid c,p,D,\Lambda)$. This approximation assumes that the intersections $\Omega^i_1\cap \Omega^j_1$ have vanishingly small probability. This is reasonable because the event $\Omega^i\cap \Omega^j$ presumes an existence of more than one randomly selected person matching the POI genotype at all loci. Forensic identification works on the assumption that this almost never happens.
Combining \eqref{model:eqn:expandpriors} and \eqref{model:eqn:bayfaccalcsplit} we get
\begin{equation}
    \frac{B_1}{B_2}\approx\frac{1}{N}\sum_{n=1}^N\sum_{i=1}^k\prod_{l=1}^L \frac{\sum_{A^l\in\Omega_1^{i,l}} \pi^l(A^l) \int_{R^l} G^l(a)\, da}{\pi^l(S^l)\sum_{{A}^l\in\Omega_2^l} \pi^l({A}^l) \int_{R^l} {G}^l(a)\, da}.\label{model:eqn:bayfacestimator}
\end{equation}

Similar calculations show that the Bayes factor is bounded from above by
\begin{equation}
\frac{B_1}{B_2} \leq \prod_{l=1}^L \frac{1}{\pi^l(S^l)}\label{eqn:bayfacupperbd}.
\end{equation}
Indeed, in examples where the POI was contributing at a high quantity, the authors saw Bayes factors very near this level, since the vast majority of the posterior probability was concentrated in $\Omega_1$. 

In order to generate the sample $p_n,c_n$ required for \eqref{model:eqn:bayfacestimator}, a Metropolis-within-Gibbs sampler was used. The details are in the Supplementary Material.

\subsection{ Fitting a density to RFL distances}\label{model:sec:implementdeets}

Equation~\eqref{eq:likelihood} requires that the density $f$ of the RFL distance between parent sequences and artifacts needs to be specified.
The model presented up to this point is flexibly usable with different MPS-based STR typing kits. The particular properties of any given kit result in different $f$. This section provides an outline for fitting the density to the particular sequencing kit used by NIST (i.e., PowerSeq) used in generating our training data. This outline represents a pipeline usable with other kits, with the difference being in the weights of the RFL distance and parameters of the density $f$. 

This density will be fitted based on the sequencing artifact distribution observed in the data, through a first-order mathematical model of PCR. Ground-truth-known homozygous loci are used for training $f$, since otherwise the artifacts are not able to be identifiably linked to an allele. The method of comparison between alleles and artifacts is the RFL distance introduced in Section~\ref{intro:sec:stringeditdists}. The distance is based on the edit types previously mentioned: stutter, indels, and SNPs. A goal of the RFL distance was to make the relationship between frequency and distance monotonic, so rare edits of an allele are farther away from that allele than common edits. In this light, estimating frequency of edits becomes important. However, in the training data, about $10 \%$ of artifacts are more than one edit away from the allele, on average. The PCR model elaborated in the Supplement 
is a principled solution to account for those multiple-edit artifacts when estimating edit probabilities.

These estimated probabilities must then be transformed into edit costs, with the rarest edits being the most expensive. Once a density family is chosen, it is inverted for the probability-cost transformation. Parameters for the density are chosen via an iterative optimization scheme to fit a theoretical CDF to the empirical CDF of distances obtained. Stability of this optimization scheme is established by various initializations. This process is outlined in the Supplement. 

\begin{figure}
  \centering
  \includegraphics[width=\linewidth]{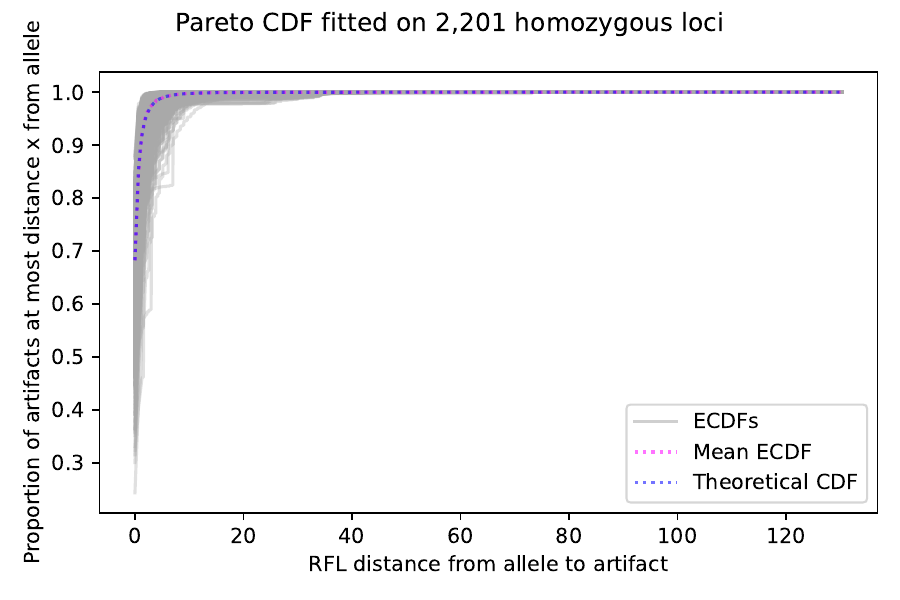}
  \caption[Fitted CDF vs. ECDFs.]{Translucent plots of the pointwise average ECDF in a pink dotted line underneath the fitted Pareto CDF in a blue dashed line, overlaid on top of 2,201 ECDFs from the individual loci.}
  \label{model:fig:fittedcdf}
\end{figure}

The fitted parameters give a Pareto CDF that closely matches the mean ECDF, shown in Figure~\ref{model:fig:fittedcdf}, overlaid on top of the individual ECDFs. Plotting of the CDFs begins at $0$ since all distances are nonnegative and the CDF is identically $0$ for lower values, and ceases after the greatest distance at any locus, exceeding $120$, at which point all CDFs are identically 1.

\section{ Results}\label{sec:sim}
This section discusses testing the Bayes factor estimator on simulated DNA mixtures to test the deconvolution algorithm.

In computing a Bayes factor using the importance sampling estimator described in Section~\ref{model:sec:bayesfac}, two models must be described, one nested inside the other. This section illustrates what happens in various scenarios. The first is a two-person mixture where the victim's DNA profile is known to be present and the POI's DNA is being tested. The second case is a two-person mixture where the victim is not present. The third case is a three-person mixture where the victim is known to be present.

It is possible, although highly unlikely as our model assigns this event a prior probability $\prod_{l=1}^L \pi^l(S^l)$, that another person from the population will randomly match the POI's DNA at all loci. The nested sub-model compares a mixture with a profile constrained to be equal to the POI's against a mixture where that profile is allowed to vary across all possible profile possibilities, including the POI's profile. 

\subsection{Synthetic Mixtures}

The data available was a collection of single-source DNA samples (see Supplementary Material for details). Because the model parameters were tuned to a particular DNA kit, it was vital that the mixtures came from that same kit. Because of funding limitations in providing true lab-mixed samples, a process was devised to synthetically mix single-source samples at desired proportions.

In order to synthetically mix files, a fixed number $k$ of contributors was determined and files $f_1,\dots,f_k$ selected, where each $f_i$ from NIST's original data is comprised of 22 known amplified and sequenced loci of a known single-source contributor. For the examples computed in this chapter, $k\in\{2,3\}$.

The majority of loci showed sequence count totals around $2500$ across all artifacts, with some variability. Thus, the synthetic mixture had counts at each locus $l$ generated IID as $N_{draw,l}\sim N(2500,200)$. In experimentation, the authors found that the overall results were very robust to variation in read depth. The work presented here does not address unique circumstances that arise due to read depth variation.

A true mixture proportion $k$-vector $p_{mix}$ was decided, corresponding to the mixing proportions of files $f_1,\dots,f_k.$ For every locus $l$, $N_{draw,l}$ sequences were present in the mixture, selected from files $f_1,\dots,f_k$ with proportions $p_{mix}$. Thus, for the $k$-vector $v_l$ a vector of counts summing to $N_{draw,l}$: $$v_l\sim \mbox{Multinomial}(N_{draw,l},p_{mix}).$$

For every locus $l$, a spread of counts $v_l$ has been generated from each file's corresponding locus $l$: $f_{1,l},\dots,f_{k,l}.$ For locus $l$ and file $i,$ $v_{l,i}$ sequences must be taken from locus $l$ in file $i.$ Label the vector of sequencing read counts in $f_{i,l}$ as $\xi_{i,l}.$ Then $\xi_{i,l}$ in $f_{i,l}$ is normalized by its sum to form $\frac{\xi_{i,l}}{\sum\xi_{i,l}},$ a selection probability vector for sequences in $f_{i,l}$ weighted by how common the sequence was in the original data. Mixture file sequences $w_{i,l}$ are then selected from file $i$ at locus $L$ in quantity $v_{l,i}$ with selection vector $\frac{\xi_{i,l}}{\sum\xi_{i,l}}$ -- i.e. $w_{i,l}\sim$ Multinomial$(v_{l,i},\frac{\xi_{i,l}}{\sum\xi_{i,l}})$.

Artifact overlap is common between files at the same loci, especially if the true parent alleles are similar. The collection of generated sequences $w_{1,l},\dots,w_{k,l}$ is concatenated across $i=1,\dots,k$ with these duplicates being merged to form the mixture at locus $l$.

\subsection{Computation}

To create mixtures to test the importance sampling estimator, five real data files out of the 661 available were selected at random to store precomputed RFL distances. For every locus $l$, the distance was computed from every true allele to every artifact across the five files.

Various factors were changed across different MCMC runs, e.g., which files were mixed, the mixing proportions of the chosen files, and the starting location for the Gibbs sampler. Two of the batches varied whether or not the POI being tested was present in the mixture (i.e., were the POI's true alleles chosen from alleles that were at high counts in the mixture, or were they chosen from alleles that were present at either an artifact level or not present at all). The combinations of these different factors followed a full factorial experimental design. The victim's mixing coordinate was varied to ensure the model performed under low and high victim contribution.

At each locus $l$, the five contributors' files could be combined to form a list of potential alleles ${K}^l$, since NIST had oracle information about the five people. In simulation, any allele in ${K}^l$ that exceeded 0.25 \% of the total artifact count at each locus ($\approx$ 2500) was placed into the list of candidate alleles $K^l$. However, whenever there was a known contributor (e.g. POI, victim), those alleles were added to $K^l$ regardless of their count in the data -- even if the count was 0. Our computer code gives a warning whenever a POI's or victim's allele appeared below the 0.25 \% of the total artifact count, since this is a sign that there is something wrong with the assertion that a victim or POI are present in the mixture. From a higher level viewpoint, if multiple loci are missing alleles that ``should'' be there, one might question the necessity of running a model at all. Indeed, such a situation results in extremely small Bayes factors -- as low as $10^{-70,000}$ in select cases. In fact, we have never encountered a single false positive in our numerical experiments.

In a variety of experiments, the Gibbs sampler proved to be highly robust to initial conditions. E.g., even with highly skewed initialization at $[0.95,0.05]$, it performed similarly and converged at a similar rate as it did with $[0.6,0.4]$ or $[0.75,0.25]$. The Gibbs sampler initialized the parameter $c$ to $22$ across all runs. Within batches, each run was given a different random seed to establish robustness of the process. The vector parameter $p$ was sorted after every step of the Gibbs sampler in descending order to maintain identifiability, with the rows of $A^l$ sorted to follow suit.

\subsubsection{Victim + POI vs. Victim + Random}\label{sim:subsec:vicsusvsvicrand}

For the MCMC runs analyzed in this section, the victim was held fixed as a known contributor in both models, and the goal was to detect the POI in a 2-mixture with proportion $p^{mix}_0$ varying over \[[0.05,0.10,0.15,0.25,0.45,0.5,0.55,0.75,0.95],\] with the victim's mixing proportion automatically equal to $1-p^{mix}_0$ in each case. The Gibbs sampler varied initialization of $p$ between $[2/3,1/3]$ and $[4/7,3/7]$, the normalization of $2:1$ and $4:3$ contributions. The POI varied between being a contributing member of the mixture (in the first coordinate, $p^{mix}_0$) and being a non-contributor. 

There were $10$ genotype combinations, nine mixing proportions, two Gibbs initializations, and one position each for the POI being in the mix or not, for a total of 360 parameter combinations. Each MCMC simulation was run for 1,100 steps each, allowing a burn-in of 100 steps (with the trace plots suggesting that this is sufficient). 
In 98 \% of the runs, no integral approximations were set to 0, and in the remaining 2 \%, never more than 0.0007 \% of integrals were. The acceptance rates for the Metropolis-within-Gibbs proposals generally varied between 20 \% and 40 \%, with the bulk around 30 \%.

\begin{figure}
  \centering
  \begin{subfigure}[t]{0.45\linewidth}
      \centering
      \includegraphics[width=\linewidth]{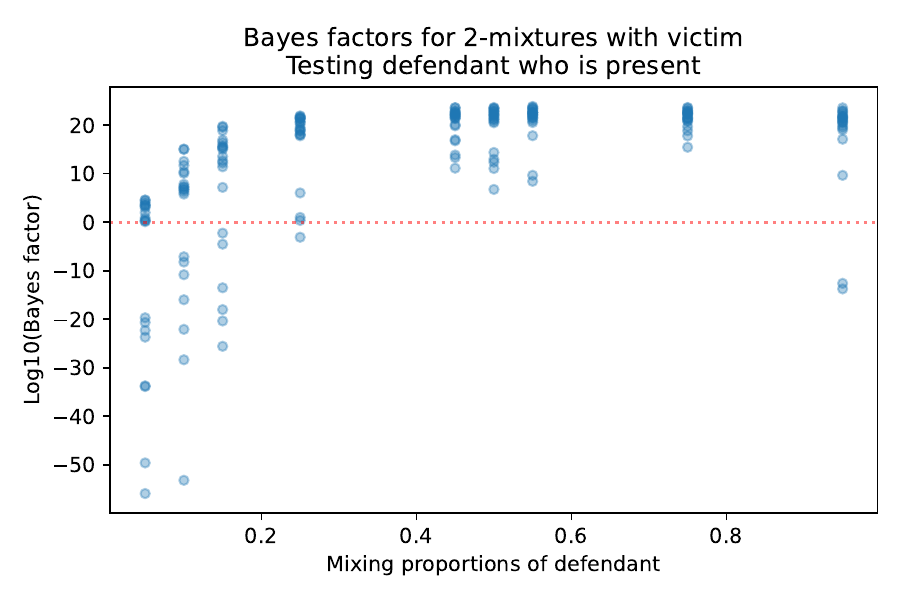}
      \caption{}
      \label{sim:fig:bayfac_vicsus2mix:subfiga}
  \end{subfigure}
  \hfill
  \begin{subfigure}[t]{0.45\linewidth}
      \centering
      \includegraphics[width=\linewidth]{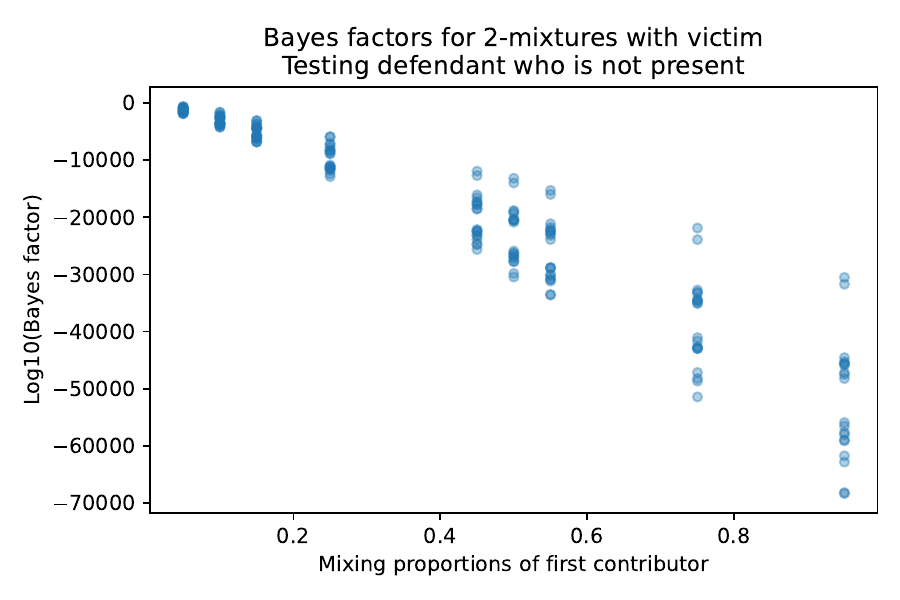}
      \caption{}
      \label{sim:fig:bayfac_vicsus2mix:subfigb}
  \end{subfigure}
  \caption[Bayes factors for victim + POI vs. victim + random.]{Bayes factors ($\log_{10}$ scale) computed over 1,000 steps after a burn-in of 100 steps for a mixture of two people, one of whom is the victim with a known profile. The red dashed line marks a Bayes factor of 0 on the log scale. In both plots the $x$-axis is the proportion of the first contributor, regardless of whether they are a POI, with the victim as the second contributor always present at proportion $1-x$. With defendant present (Figure~\ref{sim:fig:bayfac_vicsus2mix:subfiga}) the minimum log likelihood is $-55.9$, and with defendant not present (Figure~\ref{sim:fig:bayfac_vicsus2mix:subfigb}) the maximum is $-635.5$, showing clear separation.}
  \label{sim:fig:bayfac_vicsus2mix}
\end{figure}

The 360 runs returned Bayes factors shown in Figure~\ref{sim:fig:bayfac_vicsus2mix}. 
Note that because the victim is held constant in both models, the model with victim + POI has a fully determined $A^l$ for the numerator of the Bayes factor, because the victim's row is known which leaves only one row left to contain the POI's profile, which is also known. Along with greatly speeding up computation, compare this with the dip in the Bayes factor near 50-50 mixing proportions shown by the model without a victim in the next section.

\begin{figure}
    \centering
    \begin{subfigure}{0.45\linewidth}
        \centering
        \includegraphics[width=\linewidth]{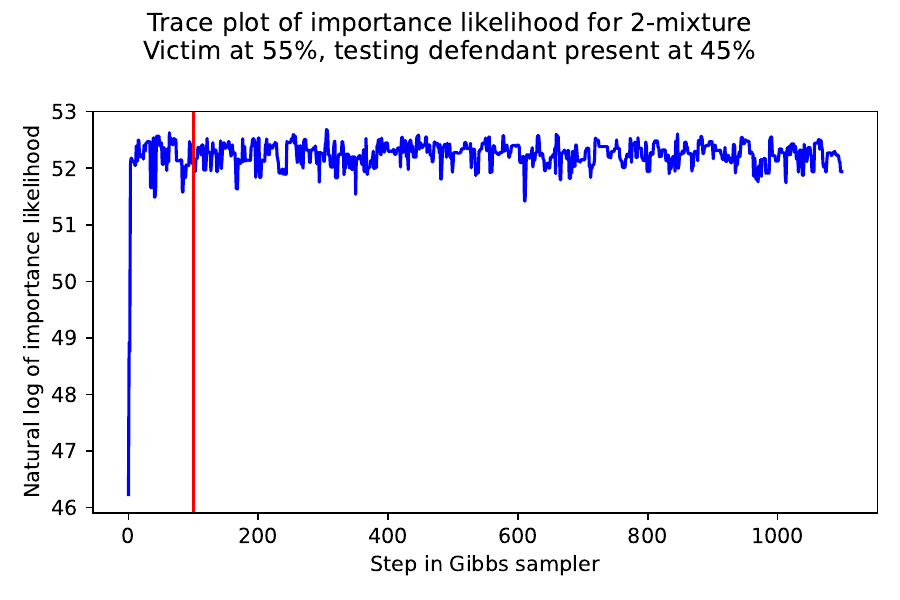}
        \caption{}
        \label{sim:fig:iliktrace_vicsus2mix:subfiga}
    \end{subfigure}
    \hfill
    \begin{subfigure}{0.45\linewidth}
        \centering
        \includegraphics[width=\linewidth]{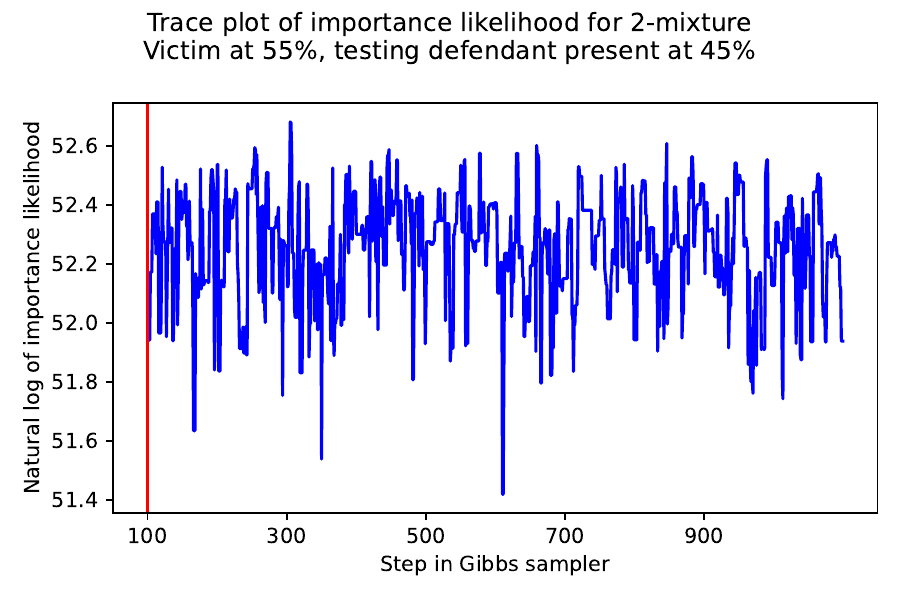}
        \caption{}
        \label{sim:fig:iliktrace_vicsus2mix:subfigb}
    \end{subfigure}
    \caption[Trace plot of Bayes factor estimator, victim + POI vs. victim + random.]{Example trace plot of the log of the stepwise importance ratio for a case where the POI was present at 45 \%. The mean of this after burn-in is the Bayes factor importance sampling estimator. Figure~\ref{sim:fig:iliktrace_vicsus2mix:subfigb} is a zoomed-in version of \ref{sim:fig:iliktrace_vicsus2mix:subfiga}.}
    \label{sim:fig:iliktrace_vicsus2mix}
\end{figure}

The Bayes factor estimate \eqref{model:eqn:bayfacestimator} is the mean of quantities that are sums of ratios. A trace plot of these quantities is shown in Figure~\ref{sim:fig:iliktrace_vicsus2mix}, showing good mixing. Example trace plots for the parameters $p$ and $c$ are shown in the Supplementary Material.

\subsubsection{POI + Random vs. Random + Random}

In this section, a POI was being tested as part of a mixture of two people where the known contributors were not present. The first mixing coordinate $p^{mix}_0$ ranged over the list \[[0.05,0.10,0.15,0.25,0.45,0.5,0.55,0.75,0.95],\] with the second coordinate $p^{mix}_1 = 1-p^{mix}_0$. For every combination of initialization $p^{mix}$ and pair of files, the parameter $p^0$ was initialized at $[0.6,0.4]$ and $[0.75,0.25]$, the normalizations of ratios $3:2$ and $3:1$. Each layout was run with the POI being a non-contributor (not present in the mixture) and with the POI being tested as the first contributor at proportion $p^{mix}_0$. 

With nine mixing coordinate vectors, 10 file combinations, two Gibbs initializations, and two possibilities for the POI being in or out of the mixture, there were 360 runs in this numerical experiment. MCMC runs were computed for 590 steps each, allowing a burn-in of 90 steps (the trace plots suggest this is sufficient). In 94 \% of the runs, no integral approximations were set to 0, and in the remaining 6 \%, never more than 0.00015 \% were. 
The acceptance rates of the Metropolis-within-Gibbs proposals generally varied between 20 \% and 40 \%, with the bulk centered around 30 \%.

\begin{figure}
  \centering
  \begin{subfigure}[t]{0.45\linewidth}
      \centering
      \includegraphics[width=\linewidth]{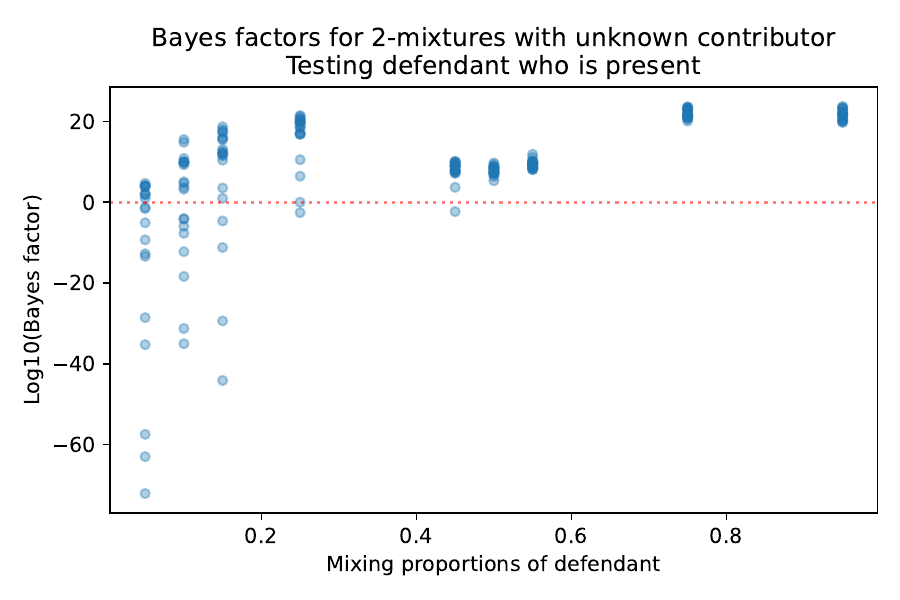}
      \caption{}
      \label{sim:fig:bayfac_randsus2mix:subfiga}
  \end{subfigure}
  \hfill
  \begin{subfigure}[t]{0.45\linewidth}
      \centering
      \includegraphics[width=\linewidth]{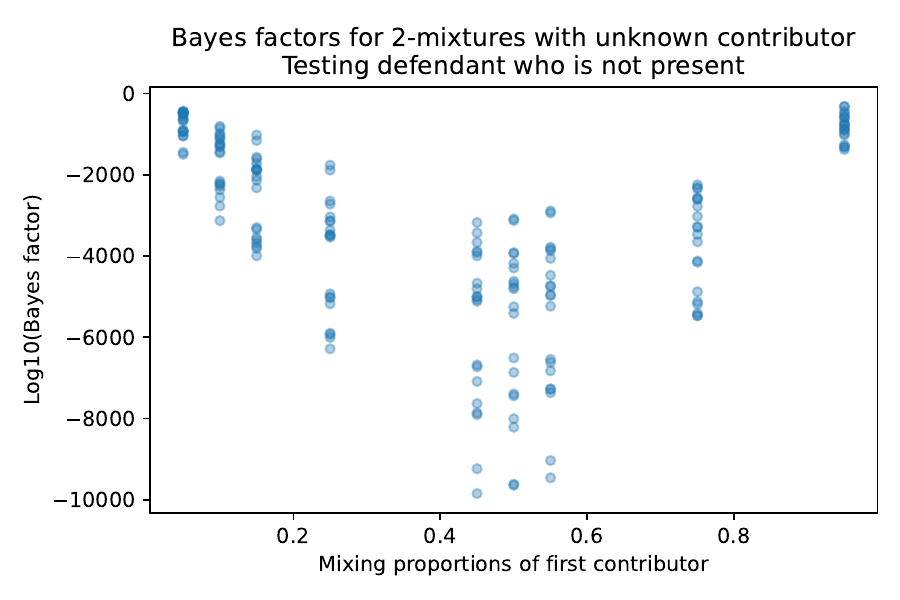}
      \caption{}
      \label{sim:fig:bayfac_randsus2mix:subfigb}
  \end{subfigure}
  \caption[Bayes factors for POI + random vs. random + random.]{Bayes factors ($\log_{10}$ scale) computed over 500 steps after a burn-in of 90 steps for a mixture of two people. The $x$-axis is the proportion of the first contributor, whether or not they are a POI, with the second contributor always present at proportion $1-x$. Note the dip in Bayes factors near 50-50 mixing proportions. When the true contributing alleles are all present at similar levels, we hypothesize there are more ways to assign people to alleles, reducing the model's certainty. With defendant present (Figure~\ref{sim:fig:bayfac_randsus2mix:subfiga}) the minimum log likelihood is $-72.2$, and with defendant not present (Figure~\ref{sim:fig:bayfac_randsus2mix:subfigb}) the maximum is $-320.8$, showing clear separation.}
  \label{sim:fig:bayfac_randsus2mix}
\end{figure}

The Bayes factors from this set of runs shown in Figure~\ref{sim:fig:bayfac_randsus2mix}. Note the dip near 50-50 mixing proportions. If the read counts of the alleles are present at a clear imbalance, the model will likely assign people either two high-count or two low-count alleles, and it is unlikely to assign alleles at drastically different proportions. Further, the vector $p$ then drives the model to give a particular person similar mixing proportions across all loci -- either high or low -- and that then affects into allele frequency. However, near $50-50$, that information is no longer useful, as everyone will have roughly equal contribution to the mixture, so the information from $p$ no longer helps discriminate between different person-allele assignments.

\begin{figure}
    \centering
    \begin{subfigure}{0.45\linewidth}
        \centering
        \includegraphics[width=\linewidth]{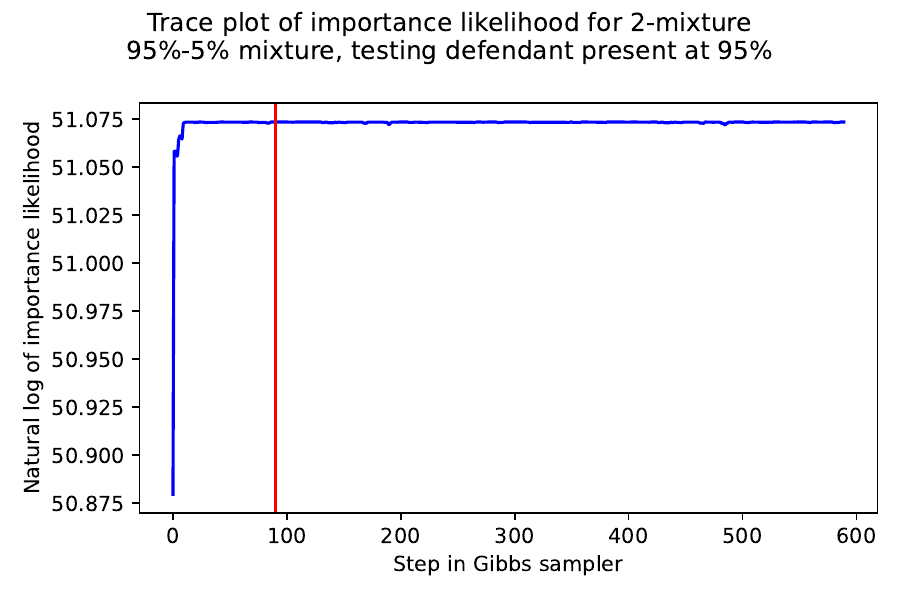}
        \caption{}
        \label{sim:fig:iliktrace_unknownsus2mix:subfiga}
    \end{subfigure}
    \hfill
    \begin{subfigure}{0.45\linewidth}
        \centering
        \includegraphics[width=\linewidth]{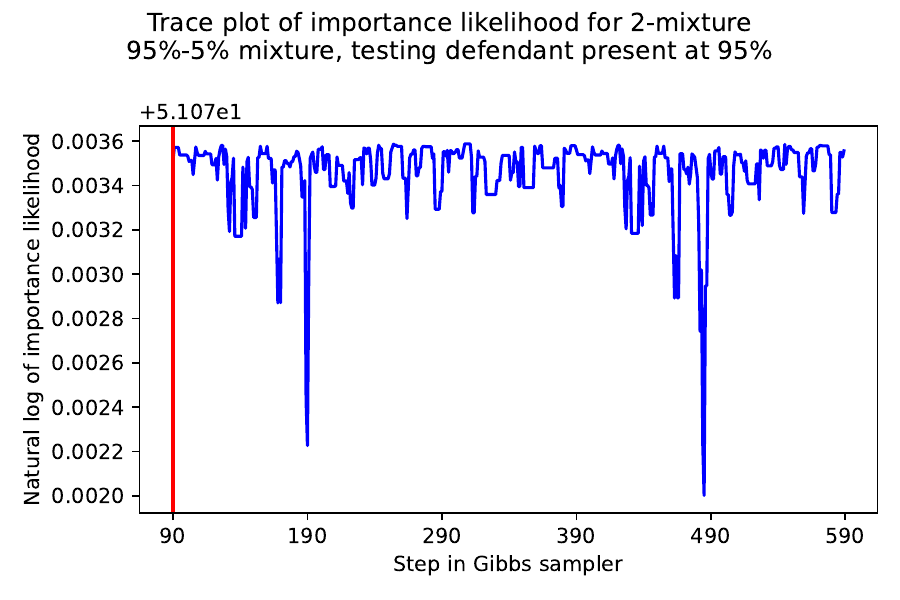}
        \caption{}
        \label{sim:fig:iliktrace_unknownsus2mix:subfigb}
    \end{subfigure}
    \caption[Trace plots of Bayes factor estimator, POI + random vs. random + random.]{Example trace plot of the log of the stepwise importance ratio, the average of which after burn-in is the Bayes factor estimator. With the POI at 95 \%, the importance likelihood approaches its maximum. The Bayes factor upper bound is discussed in Section~\ref{sec:discussion}. Figure~\ref{sim:fig:iliktrace_unknownsus2mix:subfigb} is a zoomed-in version of Figure~\ref{sim:fig:iliktrace_unknownsus2mix:subfiga}.}   
    \label{sim:fig:iliktrace_unknownsus2mix}
\end{figure}

A trace plot of the Bayes factor estimator terms as it develops across MCMC iterates is shown in Figure~\ref{sim:fig:iliktrace_unknownsus2mix}, demonstrating the Bayes factor's upper bound \eqref{eqn:bayfacupperbd}. Example trace plots for the parameters $p$ and $c$ are shown in the Supplementary Material.

\subsubsection{Victim + POI + Random vs. Victim + Random + Random}

For this set of runs, the victim was held fixed as a known contributor in both models, and the goal was to detect the POI in a 3-mixture where the proportions $p^{mix}_0$ of the POI varied over \[[0.05,0.1,0.15,0.25,0.33,0.45,0.6],\] the proportions $p^{mix}_1$ of the victim varying over $[0.2,0.6]$ (subject to the sum of the entries of $p^{mix}$ not exceeding 1), and the unknown contributor's mixing proportion $p^{mix}_2=1-p^{mix}_0-p^{mix}_1$ in each case. The Gibbs sampler varied initialization of $p$ between a normalized $16:12:9$ and $4:2:1$ vector. For the cases where the POI was not a contributor to the mixture, the Gibbs sampler initialized $p$ at $16:12:9$ every time. Note that the MCMC algorithm generally showed little variation in final answers as long as the initial $p$ was not at the simplex boundary. 

For the defendant being present this batch was run for $10$ genotype combinations, 12 mixing proportion vectors, and two Gibbs initializations, i.e., 240 runs. We ran a single Gibbs initialization for the defendant being not present in the mixture, i.e., 120 runs, for a total of 360 parameter combinations. 
MCMC runs were computed for 350 steps each, including a burn-in of 50 steps. In 66 \% of the runs, no integral approximations were set to 0, and in the remaining 34 \%, never more than 0.001 \% of integrals were. The acceptance rates of the Metropolis-within-Gibbs for $p$ varied between 12 \% and 30 \%, with the bulk centered around 22 \%. For $c$ the acceptance rates varied between 25 \% and 45 \%, with the bulk centered around 32 \%. 

Recall from Section~\ref{sec:model} that at step $t$, the proposed $p^{t+1}$ is being pulled from Dir($\alpha p^t + \vec{\beta}$) where $\alpha=25$ is a tuning parameter. In this case, with three contributors, $\alpha=70$ to increase acceptance rates in the higher dimension.

\begin{figure}[!tb]
  \centering
  \begin{subfigure}{0.45\linewidth}
    \centering
    \includegraphics[width=\linewidth]{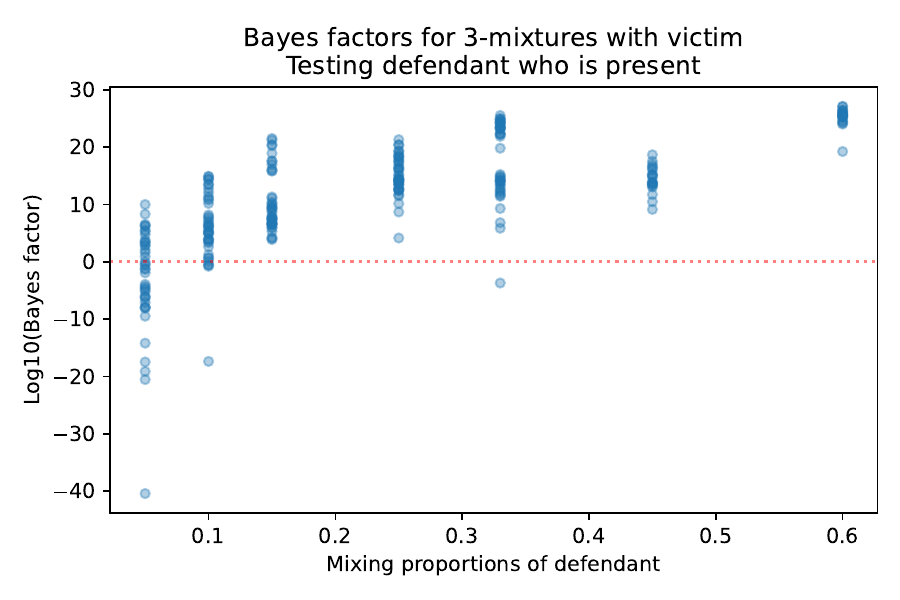}
    \caption{}
    \label{sim:fig:bayfac_vicsus3mix_susinmix}
  \end{subfigure}
  \hfill
  \begin{subfigure}{0.45\linewidth}
    \centering
    \includegraphics[width=\linewidth]{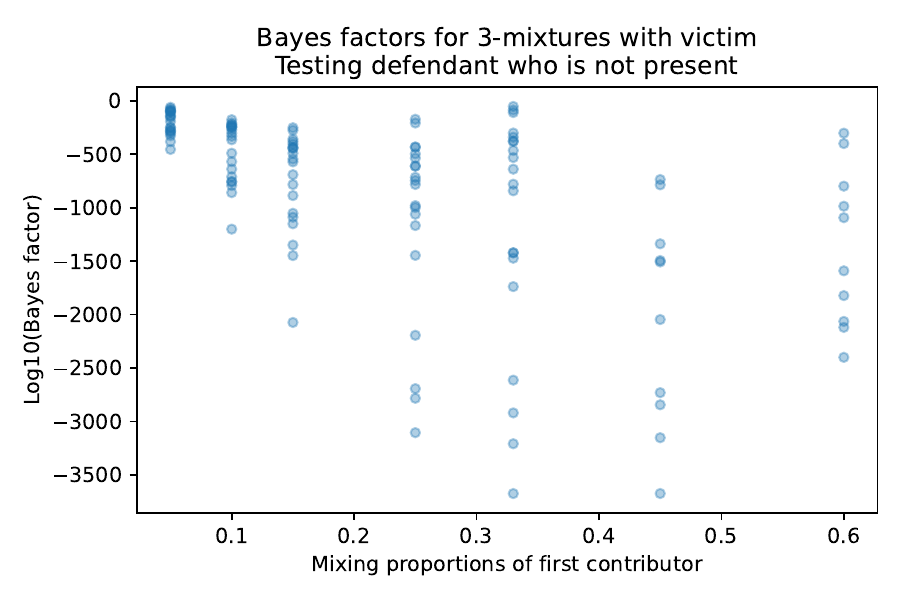}
    \caption{}
    \label{sim:fig:bayfac_vicsus3mix_ghost}
  \end{subfigure}
  \caption[Bayes factors for 3-mixture with victim.]{Bayes factors ($\log_{10}$ scale) computed over 300 steps after a burn-in of 50 steps for a mixture of three people with victim present. The $x$-axis is the proportion of the POI, who always contributes to the mixture. The victim is present at either 20 \% or 60 \%. With defendant present (Figure~\ref{sim:fig:bayfac_vicsus3mix_susinmix}) the minimum log likelihood is $-40.4$, and with defendant not present (Figure~\ref{sim:fig:bayfac_vicsus3mix_ghost}) the maximum is $-51.8$, showing clear separation.}

  \label{sim:fig:bayfac_vicsus3mix}
\end{figure}

These runs returned Bayes factors shown in Figure~\ref{sim:fig:bayfac_vicsus3mix}. POI detection is visible at all mixing proportions in Figure~\ref{sim:fig:bayfac_vicsus3mix_susinmix}, with rejection shown in Figure~\ref{sim:fig:bayfac_vicsus3mix_ghost}. Clear separation is shown in the log likelihood between cases where the defendant is present or not, even in low defendant proportions. Higher proportions of the mixture being from the defendant accentuate this pattern.

A trace plot of the Bayes factor estimator terms as it develops across MCMC iterates is shown in Figure~\ref{sim:fig:iliktrace_vicsus3mix}. Example trace plots for the parameters $p$ and $c$ are shown in the Supplementary Material.

\begin{figure}
    \centering
    \begin{subfigure}{0.45\linewidth}
        \centering
        \includegraphics[width=\linewidth]{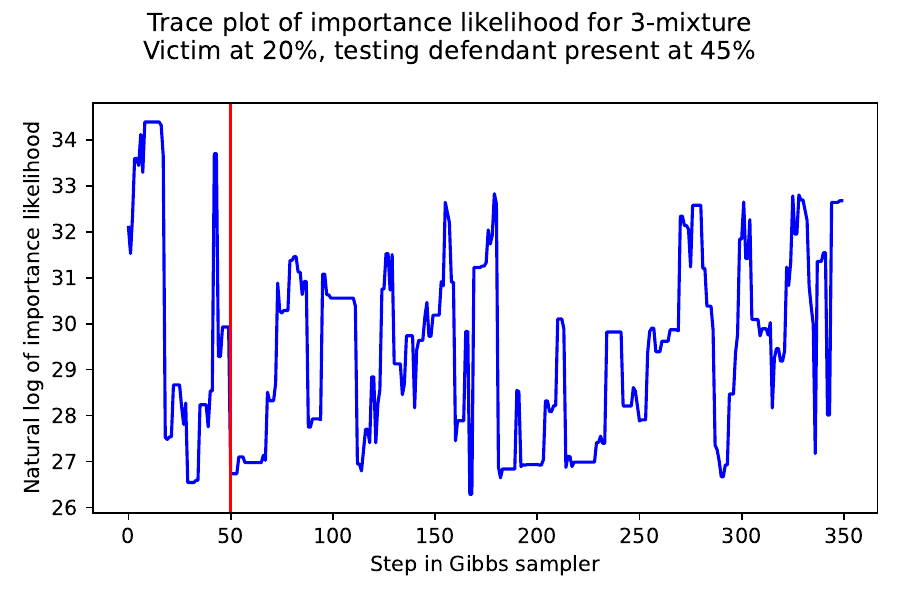}
        \caption{POI present in mixture.}
    \end{subfigure}
    \hfill
    \begin{subfigure}{0.45\linewidth}
        \centering
        \includegraphics[width=\linewidth]{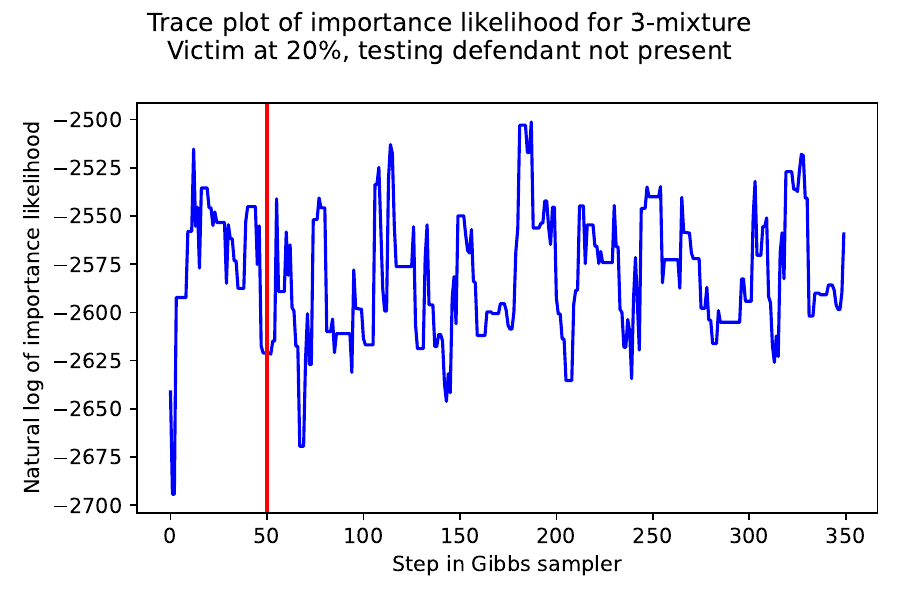}
        \caption{POI not present in mixture.}
    \end{subfigure}
    \caption[Trace plots of Bayes factor estimator, 3-mixture with victim.]{Example trace plots of the log of the stepwise importance ratio, the average of which after burn-in is the Bayes factor estimator.}
    \label{sim:fig:iliktrace_vicsus3mix}
\end{figure}

\section{ Discussion}\label{sec:discussion}

In over a thousand cases, the model showed clear evidence supporting the POI being in the mixture for mixtures containing a POI, and clear rejection of the POI being in the mixture for mixtures not containing a POI. Furthermore, trace plots suggest rapid convergence of the Markov chain Monte Carlo, which the authors hypothesize is due to the extensive marginalization already performed in summing over allele assignments $j_l$ and approximating the integral over the softmax Dirichlet allele parameters $a^l$.

Letting the allele proportions vary as a Dirichlet resulted in significant performance improvement. The sequence data does not produce perfectly balanced allele proportions, but if allele proportions are precisely equal to $A^T p$ then a perfect $50-50$ allele split of each true heterozygous locus is expected. Since significant imbalance of allele read counts was so common, allowing the likelihood to spread out around a mean of $A^T p$ prevented the model from concentrating its likelihood too tightly on particular matrices $A^l$.

It may be desirable to compute a Bayes factor comparing models that are not nested. By computing two carefully-chosen nested models, the non-nested Bayes factor can be computed if both models are part of the same larger model. One relevant example is comparing two different potential POIs (assume the victim is present for illustration). The model in this work estimates the Bayes factors $B_i/B$ of POI $i$ + victim vs. random + victim, where $B_i$ is the marginal likelihood estimate of POI $i$ + victim and $B$ is the estimated marginal likelihood of the random + victim model. Then the Bayes factor estimate for POI 1 + victim vs. POI 2 + victim is $(B_1/B)/(B_2/B) = B_1/B_2$.

The work presented in this paper also shows promise in the RFL distance being used for string similarity in forensic applications. The RFL algorithm solution proposed here fit the data as was originally desired, smoothing out the spikes in distance frequency where the regular Levenshtein distance measured stutter.

\section{ Conclusion and future work}\label{sec:conclusion}

Hundreds of forensic files were analyzed to create a table mapping loci to their highest-stuttering primary, secondary, and tertiary motifs. Following the construction of the RFL algorithm and the motif table, fitting costs with the RFL distance gave an error distribution of the PCR and sequencing process for forensic loci. Folding this string similarity density into a Bayesian data-generating model and testing it on various mixing proportions showed consistently strong ability to return large Bayes factors when the POI is present at even very low quantities, and very low Bayes factors when the POI was not present.

Future refinements are possible in several directions. Alternative dimension reduction is possible using nonparametric Bayes to fit the density of the base Levenshtein distance to the data with unit costs. The advantage of this approach is the speed and simplicity of the Levenshtein algorithm in various languages, but the disadvantage is a more complex likelihood function on distances that may not provide a clear allele-artifact mapping, since the density is not monotonic, as well as the fixed relative costs of edits. Another desirable task is checking a mixture for the presence of a profile from a list of possible contributors, e.g. a database of past criminals. For such a problem, the backwards-compatibility of this method due to using STR markers has an advantage over newer, non-standard DNA markers, as the FBI has millions of profiles available for crimes committed for the past several decades. This would become a multiple testing problem, and ensuring a low false discovery rate would be essential. 
Population allele frequencies for MPS data would refine the prior on allele profiles, but a length-based prior could be used as an approximation in the interim, with sequence lengths picked according to available tables, then sequences could be picked uniformly conditional on length \cite{allelefreqs_STR_Moretti2016}. An important future step would be acquiring empirical mixture datasets instead of synthesized mixtures, ensuring the model and methodology carries through any potential complex sequencing events.

Extremely imbalanced mixing proportions will push any model to its limits by the inherent nature of the data at hand, since eventually the POI's true allelic signal will be entirely absorbed into the error distribution of a person contributing at a much higher quantity. The Bayes factors throughout this paper showed a tendency to become less clearly separated between models where the POI was present vs. not present, but even at $5 \%$ there was still separation. Preliminary investigation suggested detecting a contributor below $1 \%$ would be very challenging. 

The integral approximation for likelihood marginalization holds potential for a speed increase. The Laplace approximation requires an optimization step, and although each optimization is reasonably fast, there are a tremendous number of them. 
A significant step could be an algorithm that uses a subset of the matrix space. At each locus, the space contains $(J_l(J_l+1)/2)^k$ matrices. The exponential growth in $k$ causes slowdown for higher numbers of contributors, even if each integral is approximated quickly. One option would be to use a genetic algorithm to create an evolving population of a small number of matrices on the scale of dozens instead of thousands or millions.

A high-priority future direction is to filter extraneous alleles out of the mixture from the list $K_l$ of known alleles, while still ensuring the alleles that are actually present are in $K_l$. Including too many extra alleles can deflate the Bayes factors. The way we started this line of work was by thresholding the data, effectively claiming that true alleles must be present above some predetermined level to be considered as true, unless they are known from a known contributor (e.g. a victim, or a POI for the prosecutor's model $M_1$), in which case they are added by assertion.

Ultimately, the use of the Restricted Forensic Levenshtein string edit distance with a Bayes factor estimated by MCMC shows promise for detecting low-contribution POIs in MPS DNA mixtures, as well as providing convincing evidence for exonerating non-contributors. The authors hope this provides a foundation for future work in advancing rigor and fairness in criminal justice worldwide.

\section{Supplementary Material}

In the supplement, the structure of the data is elaborated in detail and the mathematical notions are expanded upon.

\section{Acknowledgments}
The authors wish to thank Drs. Peter Vallone and Katherine Gettings for providing sequence data for use in illustration of our method, and also for valuable discussions we had with them that proved to be educational on many aspects of DNA biology.

\section{Funding}
Partial support for the first author is gratefully acknowledged from the National Science Foundation (NSF), award NSF-DMS-1929298 to the Statistical and Applied Mathematical Sciences Institute.

The first and second authors were supported in part by the NSF under awards DMS-1916115, DMS-2113404, and DMS-2210388.

\section{Data Availability Statement}
Note that the data that support the findings of this study are from the National Institute of Standards and Technology, but restrictions apply to the availability of these data, which were used under license for the current study, and so are not publicly available. For inquiries, please contact Hari Iyer at hariharan.iyer@nist.gov.

\section{Human Subjects Statement}
All work has been reviewed and approved by the National Institute of Standards and Technology Research Protections Office. This study was determined to be “not human subjects research” (often referred to as research not involving human subjects) as defined in U. S. Department of Commerce Regulations, 15 CFR 27, also known as the Common Rule (45 CFR 46, Subpart A), for the Protection of Human Subjects by the NIST Human Research Protections Office and therefore not subject to oversight by the NIST Institutional Review Board.

\section{Disclaimer}
Any commercial equipment, instruments, materials, or software identified in this paper are to foster understanding only. Such identification does not imply recommendation or endorsement by the National Institute of Standards and Technology, nor does it imply that the materials, equipment, or software identified are necessarily the best available for the purpose.

\printbibliography
\pagebreak
\begin{center}
\Large Supplemental Materials:\\ Bayesian DNA Mixture Deconvolution
\end{center}
\begin{refsection}
\setcounter{equation}{0}
\setcounter{figure}{0}
\setcounter{table}{0}
\setcounter{page}{1}
\makeatletter
\renewcommand{\theequation}{S\arabic{equation}}
\renewcommand{\thefigure}{S\arabic{figure}}
\renewcommand{\thetable}{S\arabic{table}}
\singlespacing

\section*{Appendix to 1: Bayesian Decision Classifier}

Changing relevant prior probabilities changes the threshold mentioned in the main body of the paper. This is generalizable via a loss function that specifies how costly each action is, with $\kappa_{ij}$ representing the loss incurred supporting $M_i$ if the true model is $j$, assuming $\kappa_{ij}>\kappa_{jj}$ for $i\neq j$. Typically one would set $\kappa{jj}=0$, i.e. no penalty for the correct decision. The expected loss of the decision $\alpha_i$ supporting $M_i$ is known as the conditional risk:
\begin{equation}
R(\alpha_i\mid E)=\sum_{j=1}^2 \kappa_{ij}P(M_j\mid E)\label{intro:eqn:conditionalrisk}.
\end{equation}
By computing $R(\alpha_i\mid E)$ for $i=1,2$, the decision can be made to support $M_i$ based on whichever has the lowest risk, i.e. supporting Model 1 precisely when \begin{equation}
    R(\alpha_1\mid E)<R(\alpha_2\mid E).\label{intro:eqn:risklessthan}
\end{equation}

Expanding \eqref{intro:eqn:risklessthan} using \eqref{intro:eqn:conditionalrisk}, 
Model 1 is supported if and only if:
\begin{equation}
    \frac{P(E\mid M_1)}{P(E\mid M_2)} > \frac{\kappa_{12}-\kappa_{22}}{\kappa_{21}-\kappa_{11}}\times \frac{P(M_2)}{P(M_1)}.\label{intro:eqn:bayfacdecision}
\end{equation}

The Bayes factor thus classifies a mixture as either containing the POI or not. 
\section*{Appendix to 2.1: Likelihood Details}

The notation $1\!\!:\!\!L$ will denote locus-indexed collections of variables, but for legibility the superscripts and subscripts are suppressed when context makes them clear.

The data for each locus contains one or two true alleles and a multitude of artifacts at varying counts, each with their respective distances from their true parent, as shown in the introduction of the paper. However, each artifact's true parent is unknown. An intermediary variable is formed to assign each artifact in the mixture to one of the possible alleles in $K^l$. 
To this end, for every locus $l$ define an assignment function $j^l:\{1,\dots,M_l\}\to \{1,\dots,J^l\}$ that intakes an artifact index $m$ out of possible indices $1,\dots,M_l$ and returns the index $j$ of the allele list $K^l = [s_{l,1},\dots,s_{l,J^l}]$ at that locus to which the artifact $y_{l,m}$ is assigned. Note $j^l$ is conditional on allele proportions $r^l$ and the data. 

Returning to data-generating process, for an artifact index $m$, $r^l_{j^l(m)}$ is the proportion of data sequences at the locus, both artifact and allele, assigned to $j^l(m)$, i.e., the true allele that $m$ is assigned to. With $d^l_{j^l(m), m}$ notation describing the distance from allele $j^l(m)$ to artifact $m$, $f(d^l_{j^l(m), m})$ represents the likelihood of generating an artifact of $s_{l,j^l(m)}$ at distance $d^l_{j^l(m), m}$.

The final step in the data generation process was to understand the likelihood of picking a particular sequence at that given distance from that given parent allele. 
Conditional on parent allele and distance, a variety of possible sequences can be generated. Any sequence with a single SNP, regardless of where the SNP is, will be the same distance from the parent. In the model, this artifact is selected uniformly at random (conditional on locus, allele, and distance). The total number of possible artifacts is a constant $C$ depending on the allele $s_{l,j^l(m)}$ and distance $d(s_{l,j^l(m)},y_{l,m}).$ 
All the data needed to determine the constant is $l$ and $m$, and everything else follows. Thus, the number of possible artifacts at the same edit distance $d$ as the artifact $m$ from allele $j^l(m)$ at locus $l$ will be denoted as $C^l_m(d).$ Letting an edit \emph{table} denote tabulated totals of edit counts from one sequence to another, $C^l_m(d)$ specifically denotes the number of possible unique edit tables distance $d$ from the given parent. Tables are used instead of paths, since here any paths with matching edit counts will share equivalent costs, and knowing the precise order of edits is unidentifiable in general.

Several assumptions are necessary to fully define the likelihood from the conditional distributions mentioned thus far. Let $\Lambda^l$ represent the totality of sequence data at locus $l$. One model assumption is that when conditioned on parent assignment $j$ and distances $D$, sequences are independent of anything other than $j$ and $D$, so $\pi(\Lambda^l\mid j^l, D^l, r^l) = \pi(\Lambda^l\mid j^l, D^l).$ Another important assumption is that, conditional on mixing proportions (the prior of which will be specified in the next section), all loci are independent, which is why a full likelihood can be taken by simply multiplying across loci. Further it is assumed that conditional on allele, artifact edit distances are independent of allele proportions $r^l$ and only depend on $j$, so $\pi(D\mid j,r) = \pi(D\mid j)$. This is because once a given allele is picked, the artifacts are generated based on that sequence alone, not using any information about other alleles or people. The generated sequences $\Lambda$ only depend on the artifact $j$ they are generated from and the distance $D$ they are from that allele, so $\pi(\Lambda\mid j,D,r) = \pi(\Lambda\mid j, D)$. This is uniform on the space of possible artifacts at the given distance, and generates the $1/C^l_{j^l(m)}(d^l_{j^l(m),m})$ term in the likelihood, where $C^l_k(d)$ is defined in the main body of the paper as the number of distinct potential artifact sequences that are exactly distance $d$ from parent allele $k$. The number of sequences $M_l$ at a locus is directly read from the data as the number of sequences $|\Lambda^l|$. Conditional on $r^l$, assignments $j^l$ are independent. Conditional on assignment $j^l$ and locus $l$, distances are independent. Conditional on assignment $j^l$, distances $D^l$, and locus $l$, sequences $\Lambda$ are independent. Artifacts are assumed to be generated as a categorical variable, proportional to the alleles they originated from.

Using this, we can write \begin{align}
    \pi(j^l\mid r^l) &= \prod_{m=1}^{M_l} r^l_{j^l(m)}\\
    \pi(D^l\mid j^l) &= \prod_{m=1}^{M_l} f(d^l_{j^l(m),m})\\
    \pi(\Lambda^l\mid j^l, D^l) &= \prod_{m=1}^{M_l}\frac{1}{C^l_{j^l(m)}(d^l_{j^l(m),m})}\\
\end{align}

Combining everything thus far results in the following likelihood, depending on person mixing proportions $p$, allele proportions $q^l$, assignments $A^l$ of alleles to people, assignments $j^l$ of artifacts to alleles, distances $D^l$ and sequence data $\Lambda^l$:
\begin{equation}
    \pi(D,j,\Lambda\mid r) = \prod_{l=1}^L \prod_{m=1}^{M_l} \frac{r^l_{j^l(m)} f(d^l_m)}{C^l_{j^l(m)}(d^l_{j^l(m),m})}.
\end{equation}

\section*{Appendix to 2.2: Priors}
 At a given locus, there is a list $K^l$ of $J^l$ different known alleles possible from the population, and any sequences detected at locus $l$ will be considered to be true alleles if they are in $K^l$, even if they are an artifact. A mixture file has multiple potential alleles which vary by locus.
 
Each person $i$ contributes a proportion $p_i$ of the total mixture. If the person is heterozygous at that locus, the $p_i$ will be divided into halves at different alleles; if homozygous, the allele combines both $p_i/2$ portions into a total of $p_i$. The assignment matrix entry $A_{ij}$, then, represents person $i$ having allele $j$ homozygously (if $A_{ij}=1$), heterozygously (if $A_{ij}=\frac{1}{2}$), or not at all (if $A_{ij}=0$). Every person must have exactly two contributing alleles (not necessarily unique), so $\sum_{j=1}^{J^l} A_{ij}=1$ for all $i$.

Each person's two alleles are drawn independently with replacement to reflect inheritance from biological parents. If the population allele frequencies were known as a vector $\gamma$, $\pi(A^l_i)$ would be the set as product of the allele frequencies indicated by $A^l_i$: in the heterozygous case $2\gamma_x\gamma_y$ for indices $x,y$ where the row vector $A^l_i = 0.5$, and in the homozygous case $\gamma_i^2$ for index $x$ where the row vector $A^l_i = 1$. The factor of $2$ in the heterozygous case is due to the fact that we do not know which allele was inherited from which parent.
This could be modified for a DNA mixture of family members, with sufficient inheritance information, and the prior could be taken on the entire matrix $A^l$ instead of individual rows to allow for allele dependence between contributors, e.g. if they are related.

The prior is customizable based on empirical population profile frequencies. Length-based frequencies are available for CE \cite{allelefreqs_STR_Moretti2016}. Allele frequencies for MPS are still in development as use of the technology expands \cite{MPSSTR_nomenclature_Parson2016, nist1036freqs_Gettings2018}, but they were not available for the data the authors had access to. Therefore, this $\gamma$ is unknown and the prior is set as independent uniform draws with replacement from the list $K^l$, reflecting one allele from each biological parent of the person, i.e., \[\pi(A^l_i)=\begin{cases}
        1/(J^l)^2 & \text{if } 1 \in A^l_i,\\
        2/J_l^2 & \text{if } 1 \notin A^l_i.
    \end{cases}\label{model:eqn:uniformindepprior}\]

Each row of $A^l$ represents a person, and each column an allele. Some loci only have a few known true population alleles, so for any mixture there will likely be instances of multiple people contributing the same allele at a locus. This is reflected in the column sums. Each column sum varies between $0$ (nobody has the allele) and $k$ (everyone is homozygous with that allele) in increments of $\frac{1}{2}.$ When multiplied by their respective proportions we get the proportional contributions of alleles 1 through $J_l$ at locus $l$ of $q:=A^{\top} p$, noting $\sum_{j=1}^{J_l} q_j = 1$. Since there exists an $A^l$ for each locus $l$, then define $q^l:=A^{l\top} p$.

The Nelder-Mead method \cite{neldermead1965} used for the optimization step of the Laplace approximation never failed to converge. However, when the vector $p$ had a coordinate very near $0$, the multinomial logistic parametrization of $p$ had an unboundedly low coordinate. In those cases, the likelihood evaluated near 0 due to the Dirichlet density. Therefore, in the rare case where the Nelder-Mead process returns a coordinate low enough that the Hessian becomes numerically ill-conditioned, the algorithm is programmed to return a Laplace approximation of $0$, effectively ignoring that matrix $A^l$ for the likelihood computation (updating $A^l$ and computing the importance likelihood estimator). This happened extremely rarely, and its frequency of occurrence is included in the sections below. 

Note that while $q$ can be zero in some coordinates, in this work we subset to nonzero coordinates for the Dirichlet by dropping the dimensions where $q=0$. This is done because the absence of an allele is important information in a genetic profile, and this must be modeled with a positive, bounded Dirichlet likelihood. Without dropping the zero coordinates of $q$, the likelihood could collapse or explode. If all parameters are forced to be nonzero, then all alleles are modeled as present at least to some extent, which does not reflect mixtures that truly lack specific alleles (particularly if the list of candidates given to the model is large).

\begin{figure}
  \vspace{-30pt}
  \centering
  \includegraphics[width=0.48\linewidth]{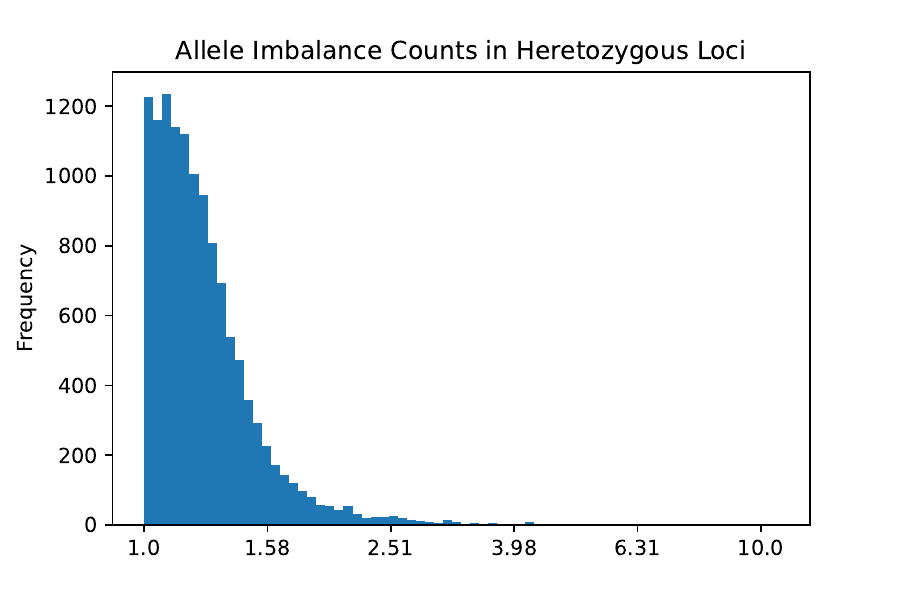}
  \vspace{-10pt}
  \caption[Heterozygous loci DoC imbalance.]{Frequencies of ratios of highest-count sequence to second-highest count sequence in 12,314 loci that are known to be heterozygous. There are 31 loci where the imbalance exceeds 4.}
  \label{model:fig:zygrats}
\end{figure}

Some of the deviation from the simpler model of $r\sim$ Categorical($q^l$) is driven by the fact that the $q^l$ reflects overly-perfect theoretical allele read count balance in heterozygous loci. While allele proportions $q^l$ force every heterozygous person to split their allele counts evenly, in 12,314 real ground-truth-known loci the imbalance ratio between counts of true alleles exceeded 3:2 more than 13 \% of the time, and twice exceeded 10:1 (see Figure~\ref{model:fig:zygrats}). By modeling conditional allele proportions $r\sim$ Dir$(cq)$, the allele assignments concentrate on the quantity $q$ but the amount this concentration happens is controlled by a hyperparameter $c$. As $c\to\infty$, $r\to q$ in distribution, and as $c\to 0$, allele proportions $r$ become increasingly disconnected from $q$. Therefore, the parameter $c$ reflects degree of belief in the vector $q$ of allele proportion assignments.

\section*{Appendix to 2.2.1: Dirichlet Parametrization}

Although commonly referred to today as ``softmax,'' which itself is an abbreviation of the more accurate ``softargmax'' \cite{deeplearning_goodfellow2016}, the multinomial logistic function dates back to Cox in 1966 \cite{logisticregressionorigins_cramer2002}, and the closely related Boltzmann distribution was discovered in the mid-1800s \cite{boltzmann1868, BoltzmannMDPI_Sharp2015}.

Note that the Dirichlet distribution on the $(I-1)$-dimensional simplex can be viewed as a function of $p_1^{I-1}\in [0,1]$ with $\sum_{i=1}^{I-1} p_i\leq 1$. This will be transformed to a function of $a_1^{I-1}\in\mathbb{R}$. For $(I-1)$-vectors $p$ and $a$, with $\alpha$ and $u_1^I$ fixed, define:
\[
    f_{Dir}(p) := \frac{\prod_{i=1}^{I-1} p_i^{u_i-1}\left(1-\sum_{i=1}^{I-1} p_i\right)^{u_I-1}}{Z_{Dir}(u)},\quad
    \textbf{p}(a) := \frac{(e^{a_1},\dots,e^{a_{I-1}})}{\sum_{i=1}^{I-1} e^{a_i} + e^{\alpha-\sum_{i=1}^{I-1} a_i}};
\]
consequently
\begin{align}
1-\sum_{i=1}^{I-1} \textbf{p}_i(a) &= 1-\frac{\sum_{i=1}^{I-1} e^{a_i}}{\sum_{i=1}^{I-1} e^{a_i}+e^{\alpha-\sum_{i-1}^{I-1} a_i}}\\
    &= \frac{\sum_{i=1}^{I-1} e^{a_i} + e^{\alpha-\sum_{i-1}^{I-1} a_i} - \sum_{i=1}^{I-1} e^{a_i}}{\sum_{i=1}^{I-1} e^{a_i} + e^{\alpha-\sum_{i=1}^{I-1} a_i}}\\
    &= \frac{e^{\alpha-\sum_{i-1}^{I-1} a_i}}{\sum_{i=1}^{I-1} e^{a_i} + e^{\alpha-\sum_{i=1}^{I-1} a_i}}\\
    &=: \mathbf{p}_I(a).
\end{align}
This function transforms $I$ nonnegative variables $p$ summing to 1 to $I$ real variables summing to $\alpha$. The Jacobian determinant of this transformation will be used in completing the change of variables:
\begin{align}
    \frac{\partial \mathbf{p}_j}{\partial a_k} &= \frac{\mathbf{1}(j=k) e^{a_j}\left(\sum_{i=1}^{I-1} e^{a_i} + e^{\alpha-\sum_{i=1}^{I-1} a_i}\right) - e^{a_j}\left(e^{a_k}+e^{\alpha-\sum_{i=1}^{I-1} a_i}(-1)\right)}{\left(\sum_{i=1}^{I-1} e^{a_i} + e^{\alpha-\sum_{i=1}^{I-1} a_i}\right)^2}\\
    &= \frac{e^{a_j}\mathbf{1}(j=k)}{\sum_{i=1}^{I-1} e^{a_i} + e^{\alpha-\sum_{i=1}^{I-1} a_i}} + \frac{e^{a_j}e^{\alpha-\sum_{i=1}^{I-1} a_i}-e^{a_j} e^{a_k}}{\left(\sum_{i=1}^{I-1} e^{a_i} + e^{\alpha-\sum_{i=1}^{I-1} a_i}\right)^2} \\
    &= \mathbf{p}_j\mathbf{1}(j=k) + \mathbf{p}_j\mathbf{p}_I - \mathbf{p}_j\mathbf{p}_k\\
    &= \mathbf{p}_j(\mathbf{1}(j=k) - (\mathbf{p}_k-\mathbf{p}_I)).
\end{align}

From this point we follow MacKay's proof in the appendix of \cite{softmaxdirichlet_mackay1998} to get $\det J = I\prod_{i=1}^I \mathbf{p}_i$, recalling $\mathbf{p}_I=1-\sum_{i=1}^{I-1}\mathbf{p}_i$. Then by standard random variable transformation:
\begin{align}
    f_A(a) &= f_{Dir}(\textbf{p}(a))|J_{\textbf{p}(a)}|\\
    &= \frac{\prod_{i=1}^{I-1} \textbf{p}_i(a)^{u_i-1}\left(1-\sum_{i=1}^{I-1} \textbf{p}_i(a)\right)^{u_I-1}}{Z_{Dir}(u)} \left(\prod_{i=1}^{I-1} \textbf{p}_i(a)\right)\left(1-\sum_{i=1}^{I-1} \textbf{p}_i(a)\right)I\\
    &= \frac{\prod_{i=1}^{I-1} \textbf{p}_i(a)^{u_i}\left(1-\sum_{i=1}^{I-1} \textbf{p}_i(a)\right)^{u_I}}{Z_{Dir}(u)/I},
\end{align}
implying the normalizing constant $Z_S(u)$ for the multinomial logistic parametrization is equal to $Z_{Dir}(u)/I$. 

Because this proof holds for any $\alpha$ (where MacKay's $a_I$ can be thought of equivalently as $\alpha-\sum_{i=1}^{I-1} a_i$ here), it is possible to integrate against any normalized density on $\alpha$, allowing $\sum_{i=1}^{I-1} a_i$ to be unconstrained. In this work, we build on MacKay's work by setting $\sum_{i=1}^{I-1} a_i \sim N(0,|u_{\neq 0}|)$ with variance equal to the number of nonzero elements of $u$, adjusting for the fact that the Dirichlet parameters $u=cq^l_{\neq 0}$ change dimension contextually, as mentioned previously. This dimension change is accomplished by viewing the Dirichlet as allowing parameter values of $0$, but redefining the remainder of the function to ignore those coordinates, effectively reducing its dimension. Other pieces of the expression may reference coordinate $j$ of the Dirichlet vector $\mathbf{p}(a)$ with the parameter $u_j=0$, and these are simply taken to be $0$.

\section*{Appendix to 2.3: Likelihood Marginalization}

A sum over $j_{1:L}$ is in fact a collection of $L$ nested sums, with each of the nested sums independently cycling over all possibilities of the function $j^l:\{1,\dots, M_l\}\to \{1,\dots,J^l\}.$ This sum over $j$ passes through the product as follows:

\begin{align}
    &\quad\ \pi(p,a,A^{1:L},D^{1:L},\Lambda^{1:L})\\
    &= \sum_{j_{1:L}} \pi(p,a,j_{1:L},A^{1:L},D^{1:L},\Lambda^{1:L}) \\
    &\propto \sum_{j_{1:L}}\left[\pi(p)\pi(c)\prod_{l=1}^L \left[\pi(A^l) \pi(a^l\mid A^l,p,c) \prod_{m=1}^{M_l} \frac{\mathbf{p}(a^l)_{j^l(m)} f(d^l_{j^l(m),m})}{{C^l_{j^l(m)}(d^l_{j^l(m),m})}} \right]\right]\label{model:eqn:beforepasssum} \\
    &= \pi(p)\pi(c) \prod_{l=1}^L \left(\prod_{i=1}^k \pi(A^l_i)\right) \frac{h_{q^l}(\sum a^l)\prod_t \mathbf{p}(a^l)_t^{(cq^l_{\neq 0})_t}}{Z_{Dir}(cq^l_{\neq 0}) / |q^l_{\neq 0}|} \prod_{m=1}^{M_l} \sum_{j=1}^{J_l} \frac{\mathbf{p}(a^l)_j f(d^l_{j,m})}{C^l_{j^l(m)}(d^l_{j^l(m),m})}.\label{model:eqn:passsum}
\end{align}

This passing the sum through the product in \eqref{model:eqn:passsum} is due to the fact that $\sum_{j_{1:L}}$ refers to $\sum_{j_1}\cdots\sum_{j_L}$, where each element $j_l$ ranges over all possible arrangements of vectors $M_l$ long with entries in $\{1,\dots,J_l\}$. I.e., $\sum_{j_l}$ is the same as $\sum_{j_l(1)=1}^{J_l}\sum_{j_l(2)=1}^{J_l}\cdots\sum_{j_l(M_l)=1}^{J_l}$. In general:
\begin{align}
    \prod_{i=1}^n\sum_{j=1}^{k_i} a_{i,j} &= (a_{1,1}+\dots+a_{1,k_1})(a_{2,1}+\dots+a_{2,k_2})\cdots (a_{n,1}+\dots+a_{n,k_n})\\
    &= \text{the sum of every possible } n\text{-product with one factor from each sum } \sum_{j=1}^{k_i} a_{i,j}\\
    &= \sum_{j_1=1}^{k_1}\cdots\sum_{j_n=1}^{k_n}\prod_{i=1}^n a_{i,j_i}.
\end{align}

After removing terms with no dependence on $j_l$, the transition from \eqref{model:eqn:beforepasssum} to \eqref{model:eqn:passsum} can be written as follows for some $T$, where $[x]=\{1,\dots,x\}$ for natural numbers $x$:

\begin{align}
    &\quad\ \prod_{l=1}^L\prod_{m=1}^{M_l}\sum_{j=1}^{J^l} T_l^m(j)\\
    &= \prod_{l=1}^L\prod_{m=1}^{M_l} (T_l^m(1)+\cdots+T_l^m(J^l))\\
    &= \prod_{l=1}^L(T_l^1(1)+\cdots+T_l^1(J^l))(T_l^2(1)+\cdots+T_l^2(J^l))\cdots(T_l^{M_l}(1)+\cdots+T_l^{M_l}(J^l))\\
    &= (T_1^1(1)+\cdots+T_1^1(J_1))\cdots (T_l^m(1)+\cdots+T_l^m(J^l))\cdots(T_L^{M_L}(1)+\cdots+T_L^{M_L}(J^L))\\
    &= \text{sum over all possible combinations of } j\in [J^l] \text{ of all products over } (l,m) \text{ of } T_l^m(j)\\
    &= \text{sum of } \prod_{l=1}^L\prod_{m=1}^{M_l} T_l^m(j^l(m))\text{ as } j_{1:L} \text{ varies over maps from } \prod_{l=1}^L [M_l]\text{ to } \prod_{l=1}^L [J^l]\\
    &= \sum_{j_{1:L}} \prod_{l=1}^L\prod_{m=1}^{M_l} T_l^m(j^l(m)).\label{model:eqn:endproofpasssum}
\end{align}

So we have that \eqref{model:eqn:endproofpasssum} proves \eqref{model:eqn:passsum}. The next step is to marginalize \eqref{model:eqn:passsum} over $a$. There are $J^l$ alleles and thus $J^l$ coordinates of $q^l$, but the Dirichlet drops the parameter dimension by 1 on the simplex. Furthermore, the Dirichlet is subsetted to only the nonzero coordinates of $q^l$. The integral over $a^l$ will be over $|q^l_{\neq 0}|$ copies of $\mathbb{R}$. For the computation below, define $R^l:=\mathbb{R}^{|q^l_{\neq 0}|}$, and $R^{1:L}:=R^1\times\cdots\times R^L$.

\begin{align}
    &\quad\ \pi(p,A^{1:L},c,D^{1:L},\Lambda^{1:L})\\
    &= \int_{R^{1:L}} \pi(p,a^{1:L},c,A^{1:L},D^{1:L},\Lambda^{1:L})\,da^{1:L} \\
    &= \pi(p,A^{1:L},c) \int_{R^{1:L}} \prod_{l=1}^L \pi(a^l\mid p,A^l,c)\pi(D^l,\Lambda^l\mid a^l,p,A^l,c)\,da^{1:L}\\
    &= \pi(p,A^{1:L},c) \int_{R^{1:L}} \prod_{l=1}^L \left(\frac{h_{q^l}(\sum a^l)\prod_{t=1} \mathbf{p}(a^l)_t^{c(q^l_{\neq 0})_t}}{Z_{Dir}(cq^l_{\neq 0})/|q^l_{\neq 0}|}\right)\prod_{m=1}^{M_l} \sum_{j=1}^{J^l} \frac{\mathbf{p}(a^l)_j f(d^l_{j,m})}{C^{l,j}_{d^l_{j,m}}}\, da^{1:L}\\
    &= \pi(p,A^{1:L},c) \int_{R^L}\cdots\int_{R^1} \prod_{l=1}^L G_l(a^l)\, da^1\cdots\, da^L\label{model:eqn:condenseG}\\
    &= \pi(p,A^{1:L},c) \int_{R^L} G_L(a^L)\cdots \int_{R^1} G_1(a^1)\, da^1\cdots\, da^L\label{model:eqn:integralproductsplit1}\\
    &= \pi(p)\pi(c) \left(\prod_{l=1}^L \prod_{i=1}^k \pi(A^l_i)\right) \left(\prod_{l=1}^L \int_{R^l} G_l(a^l)\, dr^l\right)\label{model:eqn:integralproductsplit2}\\
    &= \left(\prod_{i=1}^k p_i^{\alpha_i-1}\right) \frac{\exp\left(-\frac{(\ln(c)-\mu)^2}{2\sigma^2}\right)}{c\sigma\sqrt{2\pi}} \prod_{l=1}^L \left[\prod_{i=1}^k \pi(A^l_i)\right] \int_{R_l} G_l(a^l)\, da^l.\label{appdx:model:eqn:expandpriors}
\end{align}

where $G_l$ notation was introduced in Equation~\eqref{model:eqn:condenseG} to condense the expression:

\begin{align}
G_l(a^l) &:= \pi(a^l,D^l,\Lambda^l\mid p,A^l,c)\\
&=\left(\frac{h_{q^l}(\sum a^l)\prod_{t=1}^{J^l} \mathbf{p}(a^l)_t^{c(q^l_{\neq 0})_t}}{Z_{Dir}(cq^l_{\neq 0})/|q^l_{\neq 0}|}\right)\prod_{m=1}^{M_l} \sum_{j=1}^{J^l} \frac{\mathbf{p}(a^l)_j f(d^l_{j,m})}{C^{l,j}_{d^l_{j,m}}}.
\end{align}

In Equations~\eqref{model:eqn:integralproductsplit1} and \eqref{model:eqn:integralproductsplit2} we used the fact that

\begin{align}
    \int_{X_1}\int_{X_2} f_1(x_1) f_2(x_2)\,dx_1\,dx_2 &= \int_{X_1} f_1(x_1)\left[\int_{X_2} f_2(x_2)\,dx_2\right]\,dx_1 \\
    &= \left[\int_{X_2} f_2(x_2)\,dx_2\right]\left[\int_{X_1} f_1(x_1)\,dx_1\right].
\end{align}

\section*{Appendix to Section~2.4: MCMC}

The overarching structure will be a blocked Gibbs sampler. 
Two of the three conditional distributions will require Metropolis-within-Gibbs steps. The 
details are given below.

First, the conditional distribution for $p$:
\begin{equation}
    \pi(p\mid A, c,a, D, \Lambda) \propto \pi(p)\prod_{l=1}^L \int_{R^l} G^l(a^l)\, da^l =: h(p).
\end{equation}
The proposal distribution $b$ is $b(p^{t+1}\mid p^t)\sim~$Dirichlet$(\alpha p^t+\vec{\beta})$, where $\alpha=25$ is a tuning parameter and $\vec{\beta}$ is the hyperparameter vector $(1/k,\dots,1/k)$ used for the prior of $p$. The addition of $\vec{\beta}$ in $b$ was to reflect the chance of allelic drop-in and prevent early convergence towards a boundary. The acceptance probability $A(x\to y) = \min(1,r_h r_b)$ where $r_h=\frac{h(y)}{h(x)}$ and $r_b=\frac{b(x\mid y)}{b(y\mid x)}$.

The conditional distribution for $c$ follows similarly:
\begin{equation}
    \pi(c\mid p, A,a, D, \Lambda) \propto \pi(c)\prod_{l=1}^L \int_{R^l} G^l(a^l)\, da^l =: g(c).
\end{equation}

The proposal distribution is $b(c^{t+1}\mid c^t)\sim e^{\ln(c^t)+\eta Z}$ where $Z\sim$ N$(0,1)$ and $\eta=0.25$ a tuning parameter, so $b$ is log-normal. 
The acceptance probability $A(x\to y) = \min(1,r_g r_b)$ where $r_g=\frac{g(y)}{g(x)}$ and \begin{equation}
    r_b = \frac{b(x\mid y)}{b(y\mid x)} = \frac{\exp\left(-\frac{(\ln(x)-\ln(y))^2}{2\eta^2}\right)}{x\eta\sqrt{2\pi}}\cdot \frac{y\eta\sqrt{2\pi}}{\exp\left(-\frac{(\ln(y)-\ln(x))^2}{2\eta^2}\right)} = \frac{y}{x}.
\end{equation}

The third conditional density needed for the Gibbs sampler is $A\mid p,c,D,\Lambda,$ but independence of loci reduces this to the density $A^l\mid p,c,D^l,\Lambda^l$:
\begin{equation}
    \pi(A^l\mid p,c,a^{l},D^l,\Lambda^l) \propto \left(\prod_{i=1}^k \pi(A^l_i)\right)\int_{R^l} G^l(a^l)\, da^l.
\end{equation}

The normalization of this conditional density proceeds by summing over all possible $A^l$, so a pure Gibbs update is possible without a Metropolis step.
The sampler took initialized values of $c$ and $p$, then updated $A$, $p$, and $c$ in order for a prespecified number of steps.

\section*{Appendix to Section 2.5: Fitting a Density to the RFL Distance}

\subsection*{Markovian Graveyard for Edit Probabilities}\label{model:subsec:graveyard}

The proposed deconvolution model uses edit distance for dimension reduction, with low edit distances to reflect common edits and increasingly greater edit distances reflecting increasingly rare edits. The raw edit distance values are inconsequential; what matters are the ratios of the respective edit costs.

PCR has been studied through the lens of branching processes \cite{pcrbranch1}. The root elements in the branching process are copies of the true parent sequence, and every sequence doubles at each stage of PCR. These child sequences can either remain the same as their parent or undergo an edit (indel, SNP, or stutter). By the end of the process there is still a large portion that stayed equal to the parent the whole time, but there are some sequences that are one, two, or even dozens of edits away from the original parent. This variably-edited collection comprises the artifacts visible after the amplification and sequencing process.

\begin{wrapfigure}[12]{R}{0.5\linewidth}
\vspace{-35pt}
\begin{center}
\scalebox{.48}{
    \begin{tikzpicture}[->, >=stealth', auto, semithick, node distance=3cm]
    \tikzstyle{every state}=[fill=white,draw=black,thick,text=black,scale=1]
    \tikzstyle{every node}=[font=\Large]
    \node[state]    (N)                     {$N$};
    \node[state]    (I)[below of=N]         {$I$};
    \node[state]    (B)[left of=I]         {$B$};
    \node[state]    (F)[left of=B]         {$F$};
    \node[state]    (D)[right of=I]         {$D$};
    \node[state]    (S)[right of=D]         {$S$};
    \node[state]    (G)[below of=I]         {$G$};
    
    \path
    (N) edge[loop above]     node{$p_n$}         (N)
        edge                node[above]{$p_f$}         (F)
        edge                node[above, xshift = -0.5mm]{$p_b$}         (B)
        edge                node[above, xshift = -2.5mm]{$p_i$}         (I)
        edge                node[above, xshift = 0.9mm]{$p_d$}         (D)
        edge                node[above]{$p_s$}         (S)
    (F) edge[loop left]     node{$p_n$}         (F)
        edge                node[below]{$p_a$}         (G)
    (B) edge[loop left]     node{$p_n$}         (B)
        edge                node[below, xshift = -0.5mm]{$p_a$}         (G)
    (I) edge[loop left]     node{$p_n$}         (I)
        edge                node[below, xshift = -2.5mm]{$p_a$}         (G)
    (D) edge[loop left]     node{$p_n$}         (D)
        edge                node[below, xshift = 0.9mm]{$p_a$}         (G)
    (S) edge[loop left]     node{$p_n$}         (S)
        edge                node[below]{$p_a$}         (G)
    (G) edge[loop below]     node{$1$}           (G);
    \end{tikzpicture}
}
\vspace{-10pt}
\caption[Graveyard Markov chain.]{Graveyard Markov chain. $p_a = \sum_{edit\in\{f,b,i,d,s\}} p_{edit},$ and $p_n = 1-p_a.$}\label{graveyardchain}
\end{center}
\end{wrapfigure}
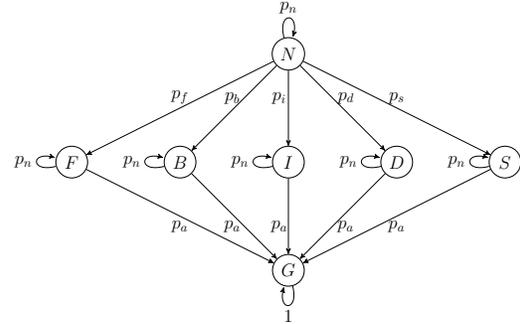

In this work the branching process is simplified for solvability. Consider a parent $P$ with five children: one each for insertion, deletion, SNPing, forward stutter of an entire motif, and backward stutter of the same, with abbreviations $i, d, s, f,$ and $b$, respectively. $P$ has one-directional arrows to each of the five children. Each of the five children has a one-directional arrow to the same kill-state, or graveyard. The graveyard represents any sequence that is two or more edits from the parent. The model forbids edits from reversing themselves, for two reasons: first, it would not be identifiable anyway, and second, the chance that a particular edit exactly reverses itself is vanishingly small. Note that the cost dictionary described in the previous paragraph is precomputed for multiple edits at once (single stutter plus multiple indels and SNPs), but here ``stutter'' strictly means one addition or removal of an exact motif. In Figure~\ref{graveyardchain}, let $p_a=p_f+p_b+p_i+p_d+p_s,$ indicating the probability of any non-null edit is the sum of probabilities of each edit.

In order to get from parent $P$ to child $S,$ where $S$ is the sequence state representing any single insertion different from the parent, the only possibility is to insert a character somewhere into the sequence $P$. Suppose this happens with probability $p_i.$ The assumption is that once an edit happens, it is negligibly likely to exactly undo itself, so no arrows return from the children to the parent. Once in the state $S,$ any additional edits to the sequence will result in it being two or more edits away from the parent, and thus in the graveyard $G.$ This means that while the transition probability from $P$ to a single-edit child $C_{edit}$ is $p_{edit},$ the transition probability from $C_{edit}$ to $G$ is $p_a:=\sum_{edits} p_{edit}.$ Once in $G,$ the sequence stays there with probability 1, due to the assumption of no edit sequence being reversible, and the further assumption that no sequence of edits will change a sequence back to being one edit from the parent. This is shown diagrammatically in Figure~\ref{graveyardchain}.

If at some point it was desirable to allow for edits to undo themselves, or sequences of edits to be performed that change a sequence to being one edit from the parent, modeling would get exponentially more complex and the edit paths to each artifact would lose identifiability, since edits could repeatedly be performed and undone ad infinitum.

To write down the transition matrix, list states 1-7 as N-FBIDS-G, corresponding to null (or equivalently ``parent''), forward, backward, insert, delete, SNP, and graveyard. This gives the transition matrix \begin{align}
    A = \begin{pmatrix}
p_n & p_f & p_b & p_i & p_d & p_s & 0\\
0 & p_n & 0 & 0 & 0 & 0 & p_a \\
0 & 0 & p_n & 0 & 0 & 0 & p_a \\
0 & 0 & 0 & p_n & 0 & 0 & p_a \\
0 & 0 & 0 & 0 & p_n & 0 & p_a \\
0 & 0 & 0 & 0 & 0 & p_n & p_a \\
0 & 0 & 0 & 0 & 0 & 0 & 1
\end{pmatrix}.
\end{align}

This matrix iterates 29 times to reflect the cycles of PCR. The proportion of sequences in each final state can be deduced directly from homozygous data, using helper functions to identify sequences in states $P,$ $C_{edit},$ and $G.$ For each homozygous locus, it is tabulated how many sequences are exactly equal to the parent, how many are exactly one deletion away, how many are exactly one SNP away, et cetera. After averaging over loci to get the post-PCR estimates, the system of equations of degree-29 polynomials can be solved to get each of the transition probabilities $p_{edit}.$

Since every sequence begins as a parent sequence (state $N$), multiply $A^{29}$ by the initial state $\mu^{(0)} = (1,0,0,0,0,0,0)$ on the left. The result is

\begin{align}
    \mu^{(29)} &= \mu^{(0)} A^{29}\\
    &=\left(p_n^{29}, 29 p_f p_n^{28}, 29 p_b p_n^{28}, 29 p_i p_n^{28}, 29 p_d p_n^{28}, 29 p_s p_n^{28}, 1-\sum_{i=1}^6 \mu_i^{(29)}\right)\\
    &= \left(p_n^{29}, 29 p_f p_n^{28}, 29 p_b p_n^{28}, 29 p_i p_n^{28}, 29 p_d p_n^{28}, 29 p_s p_n^{28}, 1-p_n^{28}(p_n+29p_a)\right),
\end{align} where the last element of the row is written as such because the row must necessarily sum to one. The number of free variables can be reduced from seven to five by using the relations

\begin{equation}
    p_n = 1-p_a=1-p_f-p_b-p_i-p_d-p_s.
\end{equation}

This system was solved as a function of the five $p_{edit}$ parameters for $edit\in\{f,b,i,d,s\}$ and using \texttt{Minimize} in Mathematica to find the least squares difference between the vector $\mu^{(29)}$ and the vector of empirical probabilities corresponding to each state. (Homozygous loci were used for this training to ensure parent-artifact identifiability.) The returned probabilities are similar but not identical to the exact solution found via \texttt{NSolve} in Mathematica, but the latter is prone to numerical instability with such high powers of small numbers.

These estimates of individual edit probabilities can be transformed into costs -- lower probability edits having a higher cost for the RFL distance, and vice versa. The probability of a SNP was higher than the probability of backward stutter. Note, however, that the probability of a given SNP -- say, a SNP at a fixed location in the string, or even more specifically, a particular letter-to-letter SNP at that location -- would be low (and thus cost would be high), but the probability of any SNP happening at any location in the string would be high (and thus cost low).

\subsection*{Converting Edit Probabilities to RFL Costs}\label{model:subsec:cdffit}

In preliminary analysis, the frequency of multiple-edit sequences was high enough to merit a Pareto for the heavy tails of the distance distribution. The distribution was zero-inflated because distance is nonnegative and there was a sizable portion (a little under 70 \%, on average) that remained equal to the parent sequence. Therefore, the following zero-inflated density was used:
\begin{equation}
    f(d\mid c,\lambda,\rho)=\rho\cdot\one(d=0) + (1-\rho)\cdot\frac{c\lambda}{(1+\lambda d)^{c+1}}\cdot\one(d>0). \label{paretopdf}
\end{equation}

The empirical probability $\hat{p}_{edit}$ of a particular edit was set equal to $f(d_{edit}\mid c,\lambda,\rho)$ and solved for the corresponding distance $d_{edit}$ --- i.e., the edit cost.

The distance cannot be zero or else that edit would have no penalty and thus unbounded numbers of those edits could be applied for free, which would defeat the purpose of the problem. Therefore $d_{edit}>0$ so we conclude the following:

\begin{equation}
    d_{edit} \propto \frac{\left(\frac{\hat{p}_{edit}}{(1-\rho)c\lambda}\right)^{\frac{-1}{c+1}}-1}{\lambda}.
\end{equation}

Following this, the single-edit distances were normalized so the cost of reverse stutter was 1.

To determine $(c,\lambda,\rho)$, the algorithm described below was seeded with various initial values and all gave the same final result. Furthermore, to preserve identifiability, this training was performed on homozygous loci, because in heterozygous data it is not knowable which parent allele is the true parent of a given artifact.

First, using the edit costs derived from the Pareto inversion with $(c_0,\lambda_0,\rho_0),$ the RFL distance of every artifact sequence to the highest-count sequence at its locus was computed (the assumption being that this will be the true parent of the artifact), and the total counts of each sequence were used to store the coordinates of a frequency-by-distance plot: $x$ as the distance, $y$ as the proportion of sequences at that locus that are distance $x$ from the parent. From these frequency-by-distance coordinates an empirical CDF (ECDF) was created via cumulative summation for each locus. The data points stored for the curve were the $x$ where actual data was obtained -- i.\,e., every point on each ECDF is a jump. Therefore, every locus had a slightly different set of $x$ coordinates, based on what distances were attained from the parent sequence at that locus. All possible attained $x$ coordinates from every locus were combined -- every distance attained from a parent to any child across any locus -- and these were combined so each ECDF had the same set of $x$ values at each possible distance. Thus, each locus now had an ECDF with jumps at each attained distance but stored values remaining flat at each distance attained at other loci but not that locus. These ECDFs were then averaged pointwise to get an overall ECDF, where the $x$-coordinates in the overall ECDF were every $x$ value attained by a sequence in the loci we were analyzing, and the $y$-coordinates in the overall ECDF were the average of every $y$-value attained at every locus. This method gave results that were superior to those obtained when using equally-spaced grid points for the $x$ coordinates of the ECDF. 

Let $\hat{F}_i$ be the ECDF for homozygous locus $i$ in index set $I$, with $n_i$ unique realized distances $(x_{ij})_{j=1}^{n_i}$, so the ECDF jumps at values $\hat{F}_i(x_{ij})$ and are flat everywhere else. Then $\hat{F}$ is the overall average ECDF with realized distances across all loci $x=\cup_{i\in I} x_i.$ For $t\in x,$ let $B_t = \{\hat{F}_i(t)\mid t\in x_i\}$. Then define $\hat{F}(t) := \frac{1}{|B_t|} \sum_{b\in B_t} b.$

A weighted $L^1$ distance was minimized between a theoretical Pareto and this average overall ECDF $\hat{F}$ by summing the absolute values of the differences of the $y$-coordinates. I.\,e., for a theoretical Pareto density in Equation~\eqref{paretopdf} there is a corresponding Pareto CDF defined for $s\geq 0$:
\begin{equation}
    F_{c,\lambda,\rho}(s) = \rho + (1-\rho)\cdot \left(1-\frac{1}{(1+\lambda s)^c}\right). \label{paretocdf}
\end{equation}

The minimum distance between the ECDF $\hat{F}$ and the CDF $F_{c,\lambda,\rho}$ was computed\footnote{ \texttt{scipy.optimize.minimize} was used with the Nelder-Mead method, max iterations 3000 with error tolerance of 1e-12, SciPy 1.8.1.} as
\begin{equation}
    \min_{c,\lambda,\rho}\sum_{t\in x} |\hat{F}(t) - F_{c,\lambda,\rho}(t)|. \label{L1minimization}
\end{equation}
This is a modified form of $L^1$ distance weighted by how isolated each point is. Highly clustered points get higher weights, while more isolated points get down-weighted. This gave better-fitting results than using a standard $L^1$ distance, albeit not dramatically different. Taking the argmin of \eqref{L1minimization} gives the answer. 
The final fitted values were $(c,\lambda,\rho) \approx (2.668, 0.513, 0.683)$, rounded to 3 decimal places.

The homozygous loci used for training were not perfectly balanced across loci. The value counts for the training loci that were available are shown in Table~\ref{model:tab:hmzygvaluecounts}.

\begin{table}
\centering
\begin{tabular}{|c|c|}
 \hline
    \textbf{Locus}  &  \textbf{Count}\\\hline
    TPOX      &  182\\\hline
    CSF1PO    &  164\\\hline
    TH01      &  158\\\hline
    D22S1045  &  145\\\hline
    D10S1248  &  139\\\hline
    D19S433   &  128\\\hline
    D2S441    &  117\\\hline
    D16S539   &  115\\\hline
    D13S317   &  102\\\hline
    D7S820    &   93\\\hline
    D8S1179   &   86\\\hline
    PENTAD    &   86\\\hline
    D5S818    &   85\\\hline
    D3S1358   &   84\\\hline
    VWA       &   83\\\hline
    D18S51    &   80\\\hline
    PENTAE    &   80\\\hline
    FGA       &   75\\\hline
    D21S11    &   59\\\hline
    D1S1656   &   53\\\hline
    D2S1338   &   52\\\hline
    D12S391   &   35\\\hline
\end{tabular}
\caption{Value counts for available homozygous loci, totaling 2,201.}\label{model:tab:hmzygvaluecounts}
\end{table}

\subsection*{Continuity Correction}

To estimate the continuity correction coefficient $C^l_k(d)$ mentioned earlier, suppose the edit path from a parent allele $s$ to child artifact $y$ is composed of $I$ insertions, $\delta$ deletions, $S$ SNPs, $F$ forward stutters, and $B$ backward stutters. The length of the parent is $LP,$ and there are $\kappa$ different motifs at the relevant locus. Then the number of possible edit tables of the same distance from the allele involves counting where the inserts, deletes, and SNPs could have happened along the length of the parent sequence via a binomial coefficient, then the number of possible types of inserts, deletes, and SNPs that can happen at each letter -- four insertions, since any letter can be added, three SNPs, since a letter can be changed to one of three others, and one deletion, since there is only one way to remove a letter. Stuttering back or forward a motif is counted differently, since losing or gaining a motif at one point in a repeating chain of motifs is unidentifiable from losing or gaining (respectively) at another point. Therefore, $$\text{ntable}(s,y) = \binom{LP}{I+\delta+S}\cdot (3^S)\cdot (4^I)\cdot (1^\delta)\cdot \binom{\kappa+F-1}{F}\cdot \binom{\kappa+B-1}{B}.$$ This quantity gives the number of possible paths between that fixed parent and child, but our estimate for the number of possible sequences observed at that distance isn't exact. It is computed by taking the sum of unique path counts over all equidistant sequences:
\begin{align}\hat{C}^{s_{l,j^l(m)}}_{d(s_{l,j^l(m)},y_{l,m})} = \sum_{\substack{m'\\ d(s_{l,j^l(m')},y_{l,m'}) = d(s_{l,j^l(m)},y_{l,m})\\ \text{ntable}(s_{l,j^l(m')},y_{l,m'}) \neq \text{ntable}(s_{l,j^l(m)},y_{l,m})}} \text{ntable}(s_{l,j^l(m')},y_{l,m'}).
\end{align}

What we have in our estimate is a lower bound of the true number (since every sequence observed was possible, but not all possible sequences were observed). The notation $C^l_m(d)$ will continue to be used, despite the fact that it's an estimate, for legibility. At the occasional locus with multiple motifs, the higher-stutter motif is used as a single-motif approximation.

\section*{Appendix to 3.2: MCMC Trace Plots}

\begin{figure}
    \centering
    \begin{subfigure}{0.45\linewidth}
        \centering
        \includegraphics[width=\linewidth]{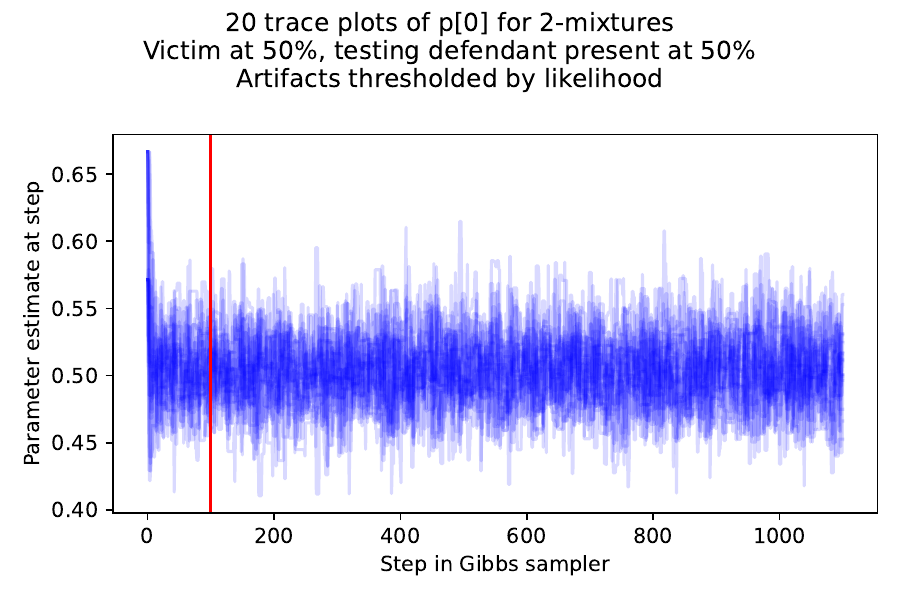}
        \caption{}
        \label{sim:fig:ptrace_vicsus2mix:subfiga}
    \end{subfigure}
    \hfill
    \begin{subfigure}{0.45\linewidth}
        \centering
        \includegraphics[width=\linewidth]{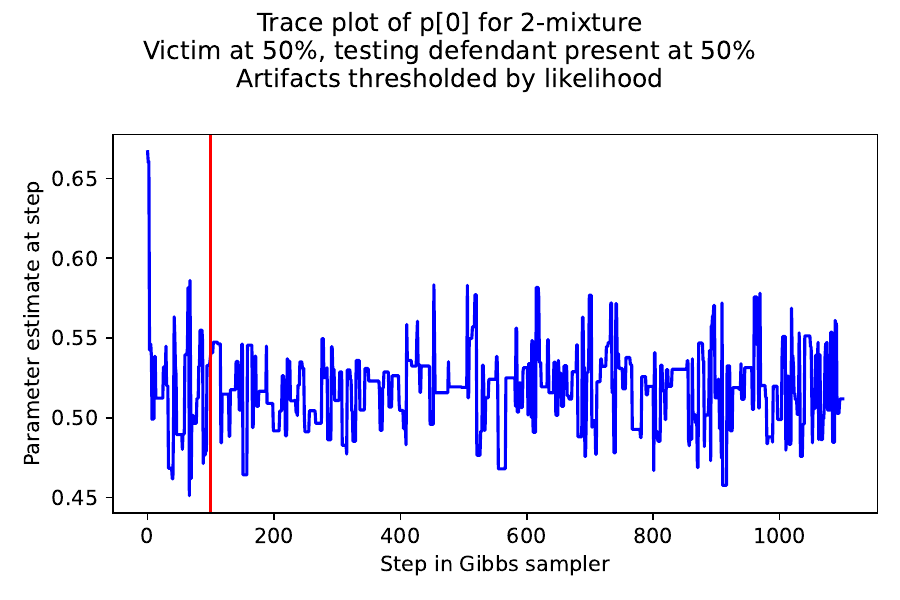}
        \caption{}
        \label{sim:fig:ptrace_vicsus2mix:subfigb}
    \end{subfigure}
    \caption[Trace plots for $p_0$, victim + POI vs. victim + random.]{Example trace plots for $p_0$, the coordinate of the POI, with $p_1=1-p_0$ by definition. The red line signifies the burn-in used. Figure~\ref{sim:fig:ptrace_vicsus2mix:subfiga} shows a batch of 20 MCMC runs with the same victim~+~POI proportions but different experimental conditions: 10 file pairing combinations with two Gibbs sampler initializations each for $p$. Figure~\ref{sim:fig:ptrace_vicsus2mix:subfigb} shows a particular member of the collection in Figure~\ref{sim:fig:ptrace_vicsus2mix:subfiga}.}
    \label{sim:fig:ptrace_vicsus2mix}
\end{figure}

\begin{figure}
    \centering
    \begin{subfigure}{0.45\linewidth}
        \centering
        \includegraphics[width=\linewidth]{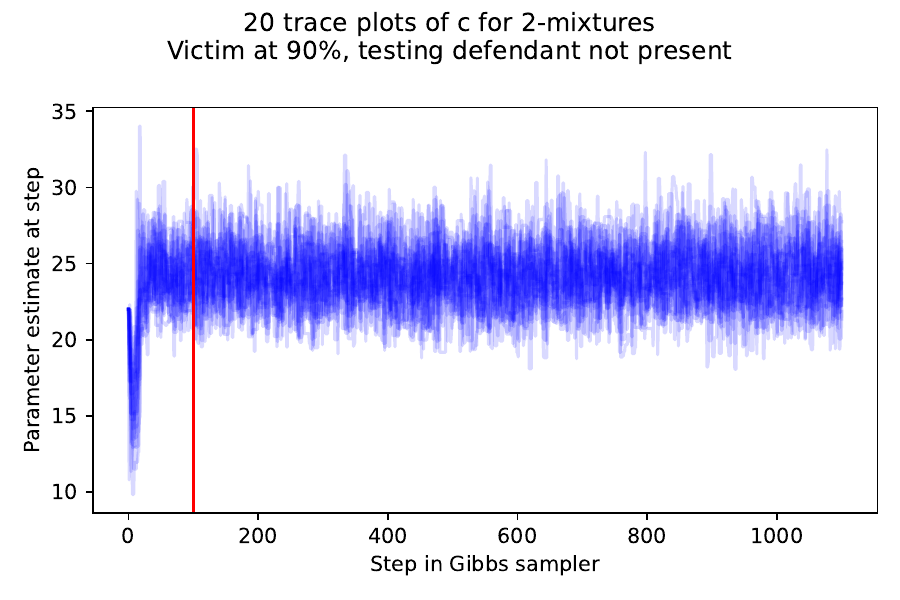}
        \caption{}
        \label{sim:fig:ctrace_vicsus2mix:subfiga}
    \end{subfigure}
    \hfill
    \begin{subfigure}{0.45\linewidth}
        \centering
        \includegraphics[width=\linewidth]{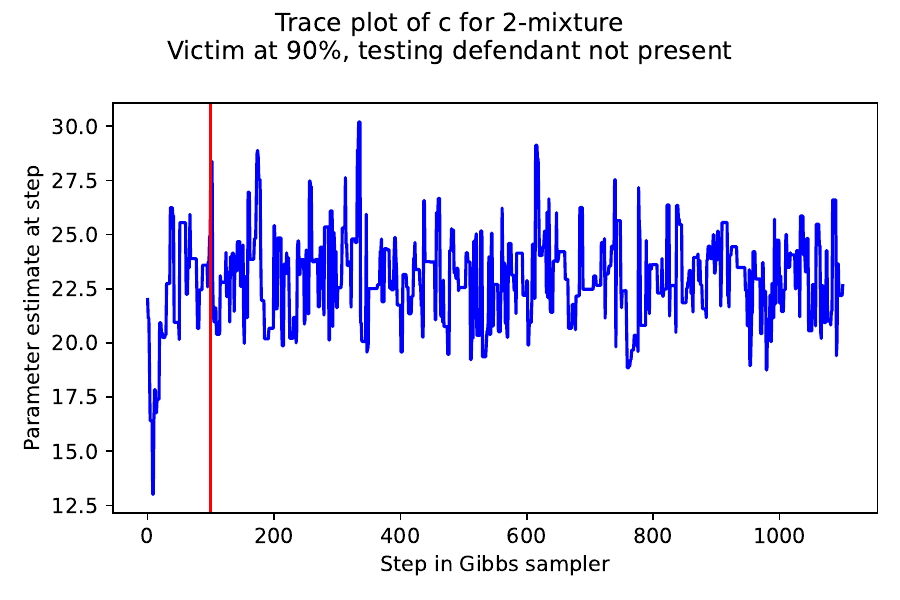}
        \caption{}
        \label{sim:fig:ctrace_vicsus2mix:subfigb}
    \end{subfigure}
    \caption[Trace plots for $c$, victim + POI vs. victim + random.]{Example trace plots for $c$ under the same victim proportions and the POI not present in the mixture under 20 different experimental conditions: 10 file pairing combinations with two Gibbs sampler initializations each for $p$. The red line signifies the burn-in period. Figure~\ref{sim:fig:ctrace_vicsus2mix:subfigb} shows a particular member of the collection in Figure~\ref{sim:fig:ctrace_vicsus2mix:subfiga}.}
    \label{sim:fig:ctrace_vicsus2mix}
\end{figure}

\begin{figure}
    \centering
    \begin{subfigure}{0.45\linewidth}
        \centering
        \includegraphics[width=\linewidth]{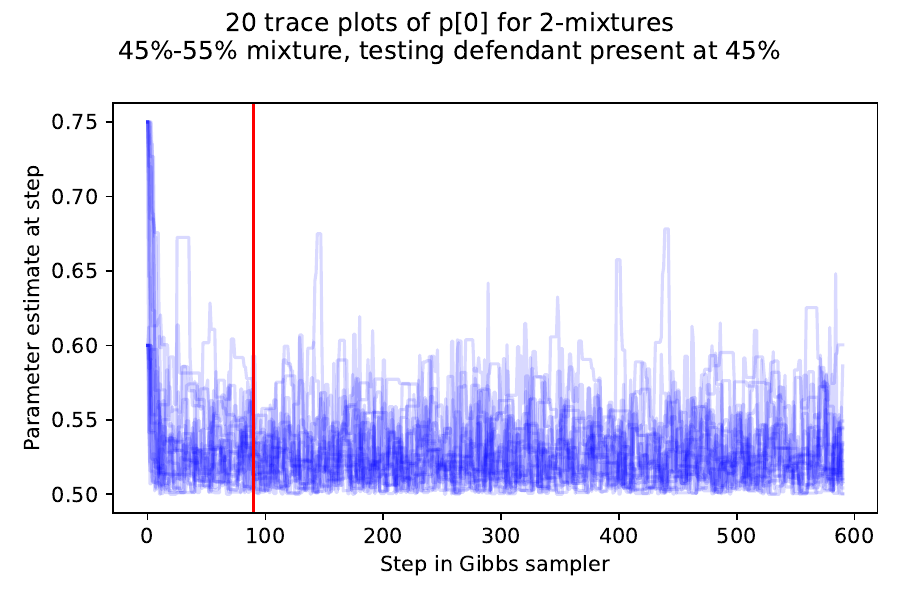}
        \caption{}
        \label{sim:fig:ptrace_unknownsus2mix:subfigwithallp}
    \end{subfigure}
    \hfill
    \begin{subfigure}{0.45\linewidth}
        \centering
        \includegraphics[width=\linewidth]{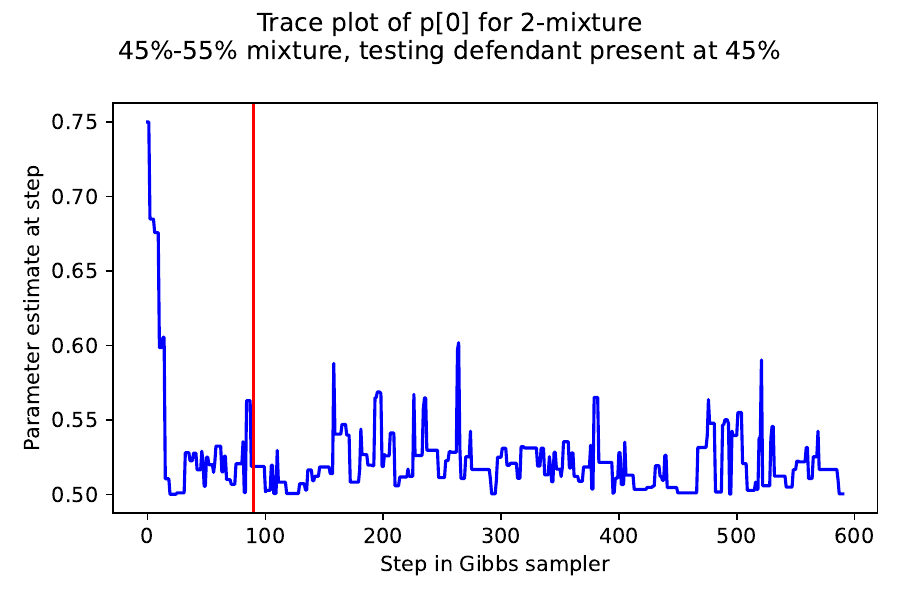}
        \caption{}
        \label{sim:fig:ptrace_unknownsus2mix:subfigb}
    \end{subfigure}
    \caption[Trace plots for $p_0$, POI + random vs. random + random.]{Example trace plots for $p_0$, the coordinate of the POI, with $p_1=1-p_0$ by definition. The red line signifies the burn-in used. Figure~\ref{sim:fig:ptrace_unknownsus2mix:subfigwithallp} shows 20 plots over 10 file mixtures with two Gibbs initializations for $p$ each. Figure~\ref{sim:fig:ptrace_unknownsus2mix:subfigb} is a single member of the collection in Figure~\ref{sim:fig:ptrace_unknownsus2mix:subfigwithallp}.}
    \label{sim:fig:ptrace_unknownsus2mix}
\end{figure}

\begin{figure}
    \centering
    \begin{subfigure}{0.45\linewidth}
        \centering
        \includegraphics[width=\linewidth]{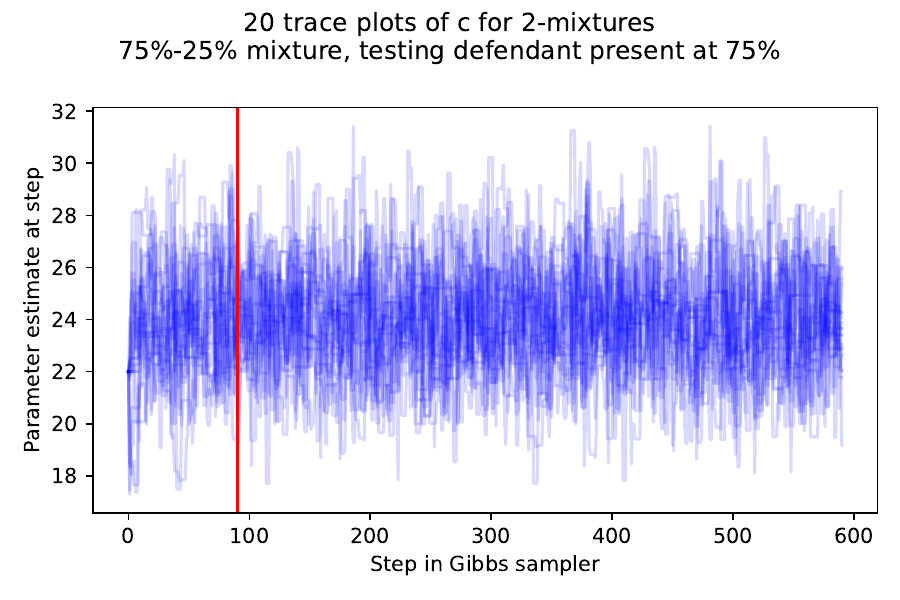}
        \caption{}
        \label{sim:fig:ctrace_unknownsus2mix:subfiga}
    \end{subfigure}
    \hfill
    \begin{subfigure}{0.45\linewidth}
        \centering
        \includegraphics[width=\linewidth]{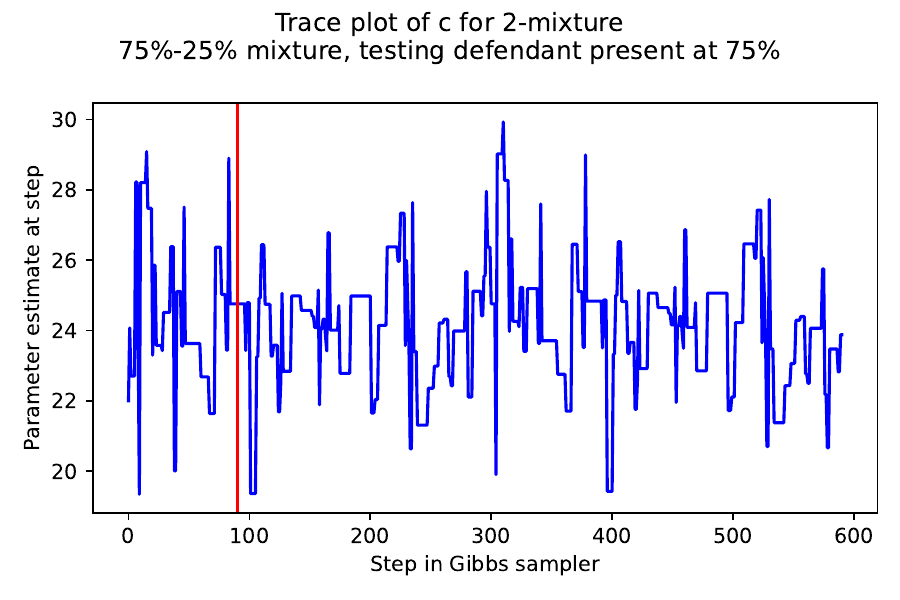}
        \caption{}
        \label{sim:fig:ctrace_unknownsus2mix:subfigb}
    \end{subfigure}
    \caption[Trace plots for $c$, POI + random vs. random + random.]{Figure~\ref{sim:fig:ctrace_unknownsus2mix:subfiga} shows a composite of trace plots for $c$ under the same victim proportions but with 20 different experimental conditions: 10 file pairing combinations with two Gibbs sampler initializations each for $p$. The red line signifies the burn-in period. Figure~\ref{sim:fig:ctrace_unknownsus2mix:subfigb} is a particular member of the collection in Figure~\ref{sim:fig:ctrace_unknownsus2mix:subfiga}.}
    \label{sim:fig:ctrace_unknownsus2mix}
\end{figure}

\begin{figure}
    \centering
    \begin{subfigure}{0.45\linewidth}
        \centering
        \includegraphics[width=\linewidth]{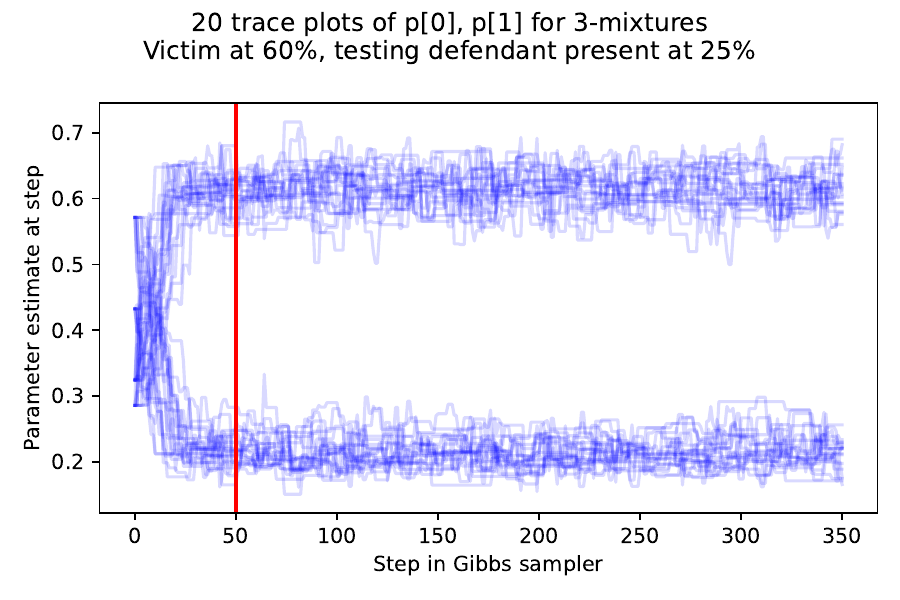}
        \caption{Superimposed trace plots for $(p_0, p_1)$.}
        \label{sim:fig:ptrace_vicsus3mix:subfiga}
    \end{subfigure}
    \hfill
    \begin{subfigure}{0.45\linewidth}
        \centering
        \includegraphics[width=\linewidth]{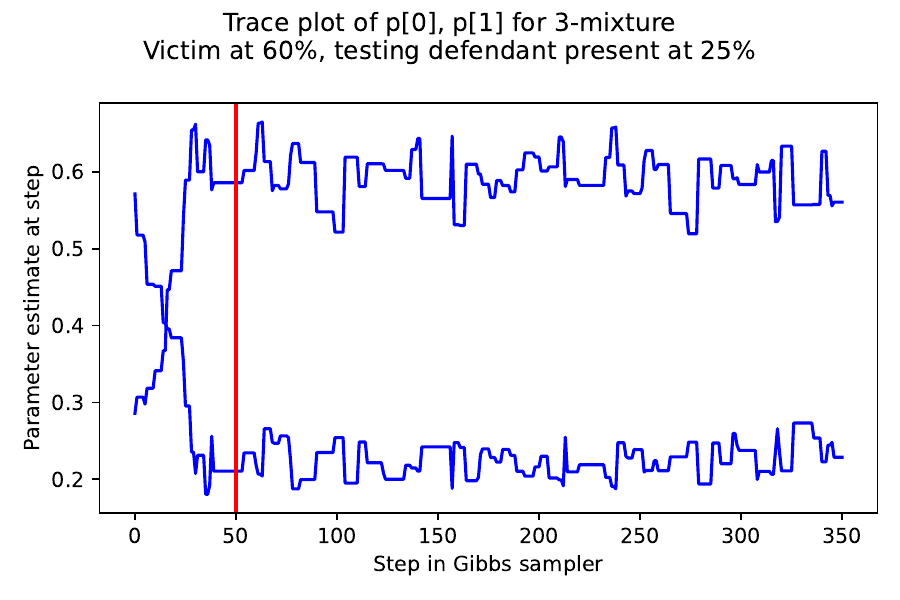}
        \vspace{-15pt}
        \caption{A single example from Figure~\ref{sim:fig:ptrace_vicsus3mix:subfiga}.}
        \label{sim:fig:ptrace_vicsus3mix:subfigb}
    \end{subfigure}
    \caption[Trace plots for $(p_0,p_1)$, 3-mixture with victim.]{Example trace plots for $p_0$ and $p_1$, with $p_2=1-p_0-p_1$ by definition.}
    \label{sim:fig:ptrace_vicsus3mix}
\end{figure}

\begin{figure}
    \centering
    \begin{subfigure}{0.45\linewidth}
        \centering
        \includegraphics[width=\linewidth]{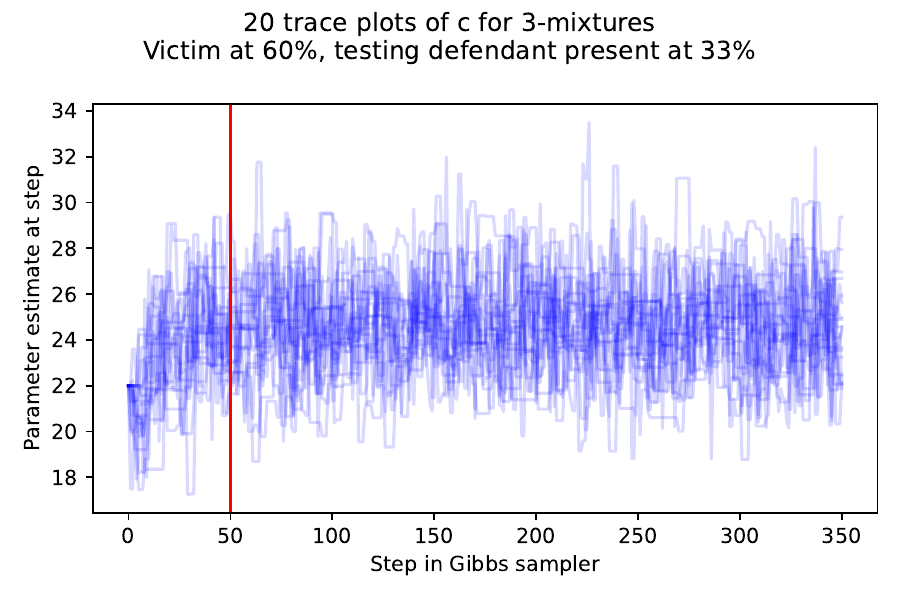}
        \vspace{-15pt}
        \caption{Superimposed trace plots for $c$.}
        \label{sim:fig:ctrace_vicsus3mix:subfiga}
    \end{subfigure}
    \hfill
    \begin{subfigure}{0.45\linewidth}
        \centering
        \includegraphics[width=\linewidth]{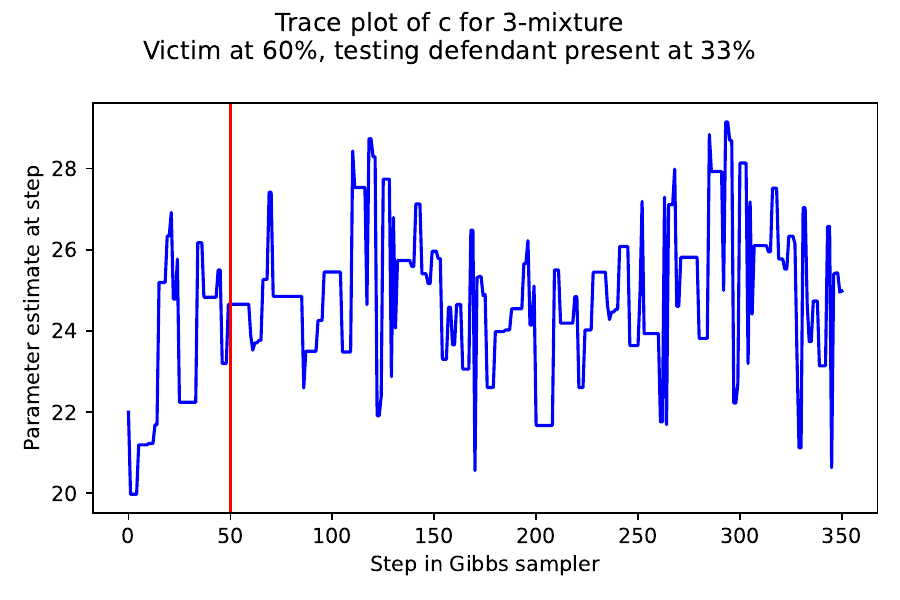}
        \vspace{-15pt}
        \caption{A single example from Figure~\ref{sim:fig:ctrace_vicsus3mix:subfiga}.}
    \end{subfigure}
    \caption[Trace plots for $c$, 3-mixture with victim.]{Composite trace plots for $c$ under the same victim proportions but with 20 different experimental conditions: 10 file pairing combinations, and two Gibbs sampler initializations for $p$. The red line signifies the burn-in period.}
    \label{sim:fig:ctrace_vicsus3mix}
\end{figure}

\clearpage
\printbibliography
\end{refsection}
\end{document}